\global\def\draftcontrol{0}

   \def\versionno{ProjCurvCosmo}

\catcode`\@=11

\expandafter\ifx\csname draftcontrol\endcsname\relax\global\def\draftcontrol{0}
\fi

{\count255=\time\divide\count255 by 60
\xdef\hourmin{\number\count255}
\multiply\count255 by-60\advance\count255 by\time
\xdef\hourmin{\hourmin:\ifnum\count255<10 0\fi\the\count255}}
\def\draftdate{\number\month/\number\day/\number\year\ \ \ \hourmin }

\newcommand\makepapertitle{\par
  \begingroup
    \renewcommand\thefootnote{\@fnsymbol\c@footnote}%
    \def\@makefnmark{\rlap{\@textsuperscript{\normalfont\@thefnmark}}}%
    \long\def\@makefntext##1{\parindent 1em\noindent
            \hb@xt@1.8em{%
                \hss\@textsuperscript{\normalfont\@thefnmark}}##1}%
     \newpage
     \global\@topnum\z@   
     \@makepapertitle
     \thispagestyle{empty}\@thanks
  \endgroup
  \setcounter{footnote}{0}%
  \global\let\thanks\relax
  \global\let\makepapertitle\relax
  \global\let\@makepapertitle\relax
  \global\let\@thanks\@empty
  \global\let\@author\@empty
  \global\let\@date\@empty
  \global\let\@title\@empty
  \global\let\title\relax
  \global\let\author\relax
  \global\let\date\relax
  \global\let\and\relax
  \def\version{\let\version\@version\@gobble}
}
\def\@makepapertitle{%
  \newpage
   \ifnum\draftcontrol=1 {}
   \version\versionno
   \vskip 3em%
   \else
   \hfill\hbox to 3cm {\parbox{4cm}{\@pubnum}\hss}%
   \vskip 3em%
   \fi
   \begin{center}%
   \let \footnote \thanks
     {\LARGE {\@title}}%
     \vskip 1.5em%
     {\normalsize
       \lineskip .5em%
       \begin{tabular}[t]{c}%
         \@author
       \end{tabular}\par}%
     \vskip 1.5em%
     {\@bstract}%
     \end{center}%
     \vskip 1.5em
     \@date%
   \par
}

\gdef\@pubnum{}
\def\pubnum#1{%
  \gdef\@pubnum{#1}}

\gdef\@bstract{}
\def\Abstract#1{%
  \gdef\@bstract{%
   \parbox{\textwidth-0pc}{%
   \centerline{\bf Abstract}\penalty1000%
\noindent
\renewcommand\baselinestretch{1.0}%
{#1}}}
}

\def\ps@paper{\let\@mkboth\@gobbletwo%
     \ifnum\draftcontrol=1
        \def\@oddfoot{\hbox to \textwidth{\tiny \versionno \hfil\tiny\draftdate}%
        \hskip -\textwidth \hbox to \textwidth{\hfil\rm\thepage\hfil}}%
     \else\def\@oddfoot{\hbox to \textwidth{\hfil\rm\thepage\hfil}}
     \fi
     \let\@evenfoot\@oddfoot
}

\def\@version#1{\ifnum\draftcontrol=1
\typeout{}\typeout{#1}\typeout{}
\vskip3mm\centerline{\hbox{\fbox{\normalsize{\tt DRAFT -- #1 -- }
                   {\draftdate}}}}\vskip3mm
\fi}
\let\version\@version
\long\def\eqlabel#1{\ifnum\draftcontrol=1
                    \tag@false  
                    \tag*{(\theequation) \hbox to -0.2cm{\hspace{0cm}\small{#1}\hss}}
                    \refstepcounter{equation}
                    \edef\@currentlabel{\theequation}
                    \ltx@label{#1}          
                    \else
                    \label{#1}
                    \fi
                    }
\let\st@bibitem\@bibitem
\let\st@lbibitem\@lbibitem
\ifnum\draftcontrol=1
  \def\@bibitem#1{%
    \st@bibitem{#1}\a@@label{#1}\ignorespaces}
  \def\@lbibitem[#1]#2{%
    \st@lbibitem[#1]{#2}\a@@label{#2}\ignorespaces}
  \def\a@@label#1{%
    \gdef\a@lab{\smash{\normalfont\small#1}}
    \ifvmode
      \if@inlabel
        \global\setbox\@labels\hbox{%
          \llap{\a@lab\let\a@lab\relax
                \kern\@totalleftmargin\kern\marginparsep}%
          \box\@labels}%
      \fi
    \fi}
\fi

\documentclass[12pt,letterpaper]{article}

\usepackage{amsmath,amssymb,array,calc,rotating,epsfig,psfrag}
\usepackage[nosort]{cite}
\usepackage{graphicx}
\usepackage{hyperref}
\usepackage{mathtools,slashed}
\usepackage[utf8]{inputenc} 
\usepackage[dvipsnames]{xcolor}

\ifnum\draftcontrol=1
\fi

\renewcommand\baselinestretch{1.25}
\renewcommand\section{\@startsection {section}{1}{\z@}%
                                   {-3.5ex \@plus -1ex \@minus -.2ex}%
                                   {2.3ex \@plus.2ex}%
                                   {\normalfont\large\bfseries}}
\renewcommand\subsection{\@startsection{subsection}{2}{\z@}%
                                   {-3.25ex\@plus -1ex \@minus -.2ex}%
                                   {1.5ex \@plus .2ex}%
                                   {\normalfont\normalsize\bfseries}}
\renewcommand\subsubsection{\@startsection{subsubsection}{3}{\z@}%
                                   {-3.25ex\@plus -1ex \@minus -.2ex}%
                                   {1.5ex \@plus .2ex}%
                                   {\normalfont\normalsize\it}}
\renewcommand\paragraph{\@startsection{paragraph}{4}{\z@}%
                                   {-3.25ex\@plus -1ex \@minus -.2ex}%
                                   {1.5ex \@plus .2ex}%
                                   {\normalfont\normalsize\bf}}

\def\revise#1       {\raisebox{-0em}{\rule{3pt}{1em}}%
                     \marginpar{\raisebox{.5em}{\vrule width3pt\
                     \vrule width0pt height 0pt depth0.5em
                     \hbox to 0cm{\hspace{0cm}{%
                     \parbox[t]{4em}{\raggedright\footnotesize{#1}}}\hss}}}}

\def\del          {\partial}

\def\tr           {\mathop{\rm Tr}}

\def\rd{{\rm d}}
\def\vevrho{<\hspace*{-3 pt}\rho\hspace*{-3 pt}>}
\def\tV{\widetilde{V}}

\def\half{{\frac12}}

\def\sqr#1#2{{\vcenter{\vbox{\hrule height.#2pt
 \hbox{\vrule width.#2pt height#1pt \kern#1pt
 \vrule width.#2pt}\hrule height.#2pt}}}}

\def\a{\alpha}
\def\b{\beta}
\def\r{\rho}

\def\O{\Omega}

\def\o{\omega}
\def\tD{{\widetilde{D}}}
\def\tcD{{\widetilde{\mathcal{D}}}}
\def\cD{{\mathcal{D}}}

\def\m{\mu}
\def\g{\gamma}
\def\l{\lambda}
\def\n{\nu}
\def\bn{\bar{\nu}}
\def\bm{\bar{\mu}}

\catcode`\@=12

\usepackage[active]{srcltx}
\usepackage{color}

\usepackage{tensind}
\tensordelimiter{?}

\begin{document}




\newcommand{\be}{\begin{equation}}
\newcommand{\ee}{\end{equation}}
\newcommand{\beq}{\begin{equation}}
\newcommand{\eeq}{\end{equation}}
\newcommand{\ba}{\begin{eqnarray}}
\newcommand{\ea}{\end{eqnarray}}
\newcommand{\nn}{\nonumber}

\def\vol{\bf vol}
\def\Vol{\bf Vol}
\def\del{{\partial}}
\def\vev#1{\left\langle #1 \right\rangle}
\def\cn{{\cal N}}
\def\co{{\cal O}}
\def\IC{{\mathbb C}}
\def\IR{{\mathbb R}}
\def\IZ{{\mathbb Z}}
\def\RP{{\bf RP}}
\def\CP{{\bf CP}}
\def\Poincare{{Poincar\'e }}
\def\tr{{\rm tr}}
\def\tp{{\tilde \Phi}}
\def\Y{{\bf Y}}
\def\te{\theta}
\def\bX{\bf{X}}

\def\TL{\hfil$\displaystyle{##}$}
\def\TR{$\displaystyle{{}##}$\hfil}
\def\TC{\hfil$\displaystyle{##}$\hfil}
\def\TT{\hbox{##}}
\def\HLINE{\noalign{\vskip1\jot}\hline\noalign{\vskip1\jot}} 
\def\seqalign#1#2{\vcenter{\openup1\jot
  \halign{\strut #1\cr #2 \cr}}}
\def\lbldef#1#2{\expandafter\gdef\csname #1\endcsname {#2}}
\def\eqn#1#2{\lbldef{#1}{(\ref{#1})}%
\begin{equation} #2 \label{#1} \end{equation}}
\def\eqalign#1{\vcenter{\openup1\jot   }}
\def\eno#1{(\ref{#1})}
\def\href#1#2{#2}
\def\half{{1 \over 2}}

\def\ads{{\it AdS}}
\def\adsp{{\it AdS}$_{p+2}$}
\def\cft{{\it CFT}}

\newcommand{\ber}{\begin{eqnarray}}
\newcommand{\eer}{\end{eqnarray}}

\newcommand{\bea}{\begin{eqnarray}}
\newcommand{\eea}{\end{eqnarray}}

\newcommand{\beqar}{\begin{eqnarray}}
\newcommand{\cN}{{\cal N}}
\newcommand{\cO}{{\cal O}}
\newcommand{\cA}{{\cal A}}
\newcommand{\cT}{{\cal T}}
\newcommand{\cF}{{\cal F}}
\newcommand{\cC}{{\cal C}}
\newcommand{\cR}{{\cal R}}
\newcommand{\cW}{{\cal W}}
\newcommand{\eeqar}{\end{eqnarray}}
\newcommand{\lm}{\lambda}\newcommand{\Lm}{\Lambda}
\newcommand{\eps}{\epsilon}


\newcommand{\nonu}{\nonumber}
\newcommand{\oh}{\displaystyle{\frac{1}{2}}}
\newcommand{\dsl}
  {\kern.06em\hbox{\raise.15ex\hbox{$/$}\kern-.56em\hbox{$\partial$}}}
\newcommand{\as}{\not\!\! A}
\newcommand{\ps}{\not\! p}
\newcommand{\ks}{\not\! k}
\newcommand{\D}{{\cal{D}}}
\newcommand{\dv}{d^2x}
\newcommand{\Z}{{\cal Z}}
\newcommand{\N}{{\cal N}}
\newcommand{\Dsl}{\not\!\! D}
\newcommand{\Bsl}{\not\!\! B}
\newcommand{\Psl}{\not\!\! P}
\newcommand{\eeqarr}{\end{eqnarray}}
\newcommand{\ZZ}{{\rm \kern 0.275em Z \kern -0.92em Z}\;}

\def\s{\sigma}
\def\a{\alpha}
\def\b{\beta}
\def\r{\backslash l}
\def\d{\delta}
\def\g{\gamma}
\def\G{\Gamma}
\def\ep{\epsilon}
\makeatletter \@addtoreset{equation}{section} \makeatother
\renewcommand{\theequation}{\thesection.\arabic{equation}}

\def\be{\begin{equation}}
\def\ee{\end{equation}}
\def\bea{\begin{eqnarray}}
\def\eea{\end{eqnarray}}
\def\m{\mu }
\def\n{\nu}
\def\g{\gamma }
\def\p{\phi}
\def\L{\Lambda,  }
\def \W{{\cal W}}
\def\bn{\bar{\nu}}
\def\bm{\bar{\mu}}
\def\bw{\bar{w}}
\def\ba{\bar{\alpha}}
\def\bb{\bar{\beta}}

\begin{titlepage}
\begin{center}
\baselineskip .9 cm
{\Large \bf Dark Energy From Dynamical Projective Connections}\\
\vskip 0.4 cm
\baselineskip .4 cm
{\bf \footnotesize  Samuel Brensinger$^{a,}$\footnote{samuel-brensinger@uiowa.edu}, Kenneth Heitritter${}^{b,}$\footnote{kenneth-heitritter@uiowa.edu }, Vincent G. J. Rodgers${^{b,}}$\footnote{vincent-rodgers@uiowa.edu}, \\
Kory Stiffler$^{c,}$\footnote{kory$_-$stiffler@brown.edu}, and Catherine A. Whiting$^{d,}$\footnote{cwhiting@bates.edu}}

\vspace{.4cm}

{\it ${}^a$ Department of Mathematics}\\
{\it  The University of Iowa,   Iowa City, IA 52242, USA}\\

\vspace{.2cm}

{\it ${}^b$ Department of Physics and Astronomy}\\
{\it   The University of Iowa,   Iowa City, IA 52242, USA}\\

\vspace{.2cm}

{\it ${}^c$ Brown Theoretical Physics Center and Department of Physics}\\
{\it Brown University, Barus Building 211, Providence, RI 02912-1843, USA }

\vspace{.2cm}

{\it ${}^d$ Department of Physics and Astronomy}\\
{\it Bates College, Lewiston, Maine 04240}

\end{center}
\begin{abstract}
We further develop the gravitational model, Thomas-Whitehead Gravity (TW Gravity), that arises when projective connections become dynamical fields.   TW Gravity has its origins in geometric actions from string theory where the TW projective connection appears as a rank two tensor, $\cD_{ab}$, on the spacetime manifold.  Using a Gauss-Bonnet (GB) action built from the \((\rd+1)\)-dimensional TW connection, and applying the tensor decomposition $\cD_{ab} = D_{ab} +  4\Lambda /(\rd(\rd-1)) g_{ab}$, we arrive at a gravitational 
model
made up of a $\rd$-dimensional Einstein-Hilbert + GB action sourced by $D_{ab}$ and with cosmological constant $\Lambda$. The $\rd=4$ action is studied and we find that  $\Lambda \propto 1/J_0$, with $J_0$ the coupling constant for $D_{ab}$. For $\Lambda$ equal to the current measured value, $J_0$ is on the order of the measured angular momentum of the observable Universe. 
We view this as $\Lambda$ controlling the scale of patches of the Universe that acquire angular momentum, with the net angular momentum of multiple patches vanishing, as required by the cosmological principle.  
We further find a universal axial scalar coupling to all fermions where the trace, \(\cD = \cD_{ab}g^{ab} \) acts as the scalar.   This suggests that \( \cD \) is   also a 
dark matter
portal for non-standard model fermions.  
\end{abstract}
\end{titlepage}
\section{Introduction}

The two great, outstanding cosmological and astrophysical problems are the natures of dark energy and dark matter. Together, they comprise roughly 95\% of the energy of the Universe, though their identities are unknown. Dark matter is hypothetically the glue holding galaxies together, since current measurements indicate that the outer regions of galaxies are spinning faster than what would be predicted from the gravitational pull of only the baryonic matter within the galaxies. Using merry-go-rounds as analogies to galaxies, the outer regions of galaxies are like children standing on the edges of rapidly spinning merry-go-rounds, with Dark matter playing the role of the adults holding on to the children to keep them from flying off.

Dark energy is the name given to the unknown substance which acts like a negative pressure, pulling the Universe apart. We are currently in an era of dark energy domination as the density of matter has become diffuse enough within the last four or five billion years for the small yet constant density of dark energy to become larger than the density of matter.
Current measurements indicate that the present action of dark energy is consistent with a cosmological constant, thus dark energy will continue to expand the universe at an accelerated rate for an infinite amount of time, given there are not changes in the identity of dark energy. 

Treating cosmological parameters as fundamental constants has a long history, though it is seldom discussed. In 1937, Dirac considered dimensionless constants involving for instance the Hubble constant $H_0$ and the charge $e$ and mass $m$ of the electron~\cite{Dirac:1937ti}. Dirac noted that $H_0 mc^3 /e^2 $  was on order of the ratio between the electric and gravitational forces between electrons and protons. In 1972, Weinberg~\cite{Weinberg:1972} reviewed this approach as an introduction to his review of Brans and Dicke's model of scalar-tensor gravity~\cite{Brans:1961sx}. More recently \cite{Alexander:2017ont} considered a relation between the vacuum energy and the hierarchy of forces.   In~\cite{Gurzadyan:2019bwx} the cosmological constant itself was considered as a fundamental constant. In this paper, we introduce the cosmic angular momentum constant $J_0$ as a coupling constant in the recently introduced tensor-tensor model of gravity based on dynamical  projective geometry~\cite{Brensinger:2017gtb}. We refer to this model as $TW$ gravity after Thomas and Whitehead's early work in projective geometry~\cite{Thomas:1925a,Thomas:1925b,Whitehead:1931a}.

While the action of $TW$ gravity will be manifestly invariant under projective transformations, it is not necessarily true that physical observables will also be projectively invariant. Indeed, cosmological observables such as the deceleration $(q)$ and jerk $(j)$ have been shown \cite{Gibbons:2010kv} to transform non-trivially under projective transformations. We also stress the point that the construction of $TW$ gravity only exploits the existence of a projective structure. It is well-known that both projective and conformal structure are necessary to fully set up the notion of space-time geometry \cite{Ehlers2012}.

We demonstrate both dark energy and some dark matter applications of $TW$ gravity. In  the context of recent work \cite{Brensinger:2017gtb}, we consider $TW$ gravity to be inspired by string theory and 2D quantum gravity. We present here the pure Gauss-Bonnet $TW$ gravity as an initial investigation. The pure GB action has the feature of becoming an Einstein-Hilbert action with an additional interaction when we use a particular ansatz for the diffeomorphism field. In particular, we are able to predict a bare  cosmological constant term in the action  that depends on an angular momentum parameter $J_0$ that we argue is of cosmological scale  (i.e., sums of galactic and/or CMB angular momenta), rather than the fundamental physics scale  $\hbar$. This predicts the  bare cosmological constant to be on the order of today's measurements. We also argue that it is more natural to take this angular momentum parameter $J_0$ of $TW$ gravity to be of cosmological scales as $TW$ gravity is a classical action taken to describe the entire Universe rather than individual particle physics experiments. We demonstrate how the Einstein-Hilbert action becomes manifest  \emph{within} this pure Gauss-Bonnet  $TW$ gravity, for a particular decomposition of the diffeomorphism field. The cosmological constant then arises naturally. This distinguishes the present strategy from other efforts where $f(R)$ gravity and/or massive gravity is included in gravitational studies of dark energy and dark matter 
\cite{Motohashi:2014una,deRham2014,Bakopoulos:2018nui,Antonio2010}.  

It is important to note that the cosmological constant we generate arises from the angular momentum parameter $J_0$ associated with a dynamical projective connection. The references\cite{Gibbons:2008ru,Nurowski:2010xc,Gibbons:2010kv} demonstrate how a projective transformation between different Friedmann-Lemaitre-Robertson-Walker metrics leads to a shift in the cosmological constant. However, they do so without association to a dynamical projective connection.

This paper is organized as follows. 
 In section~\ref{s:CCProblem} we review the cosmological constant problem and outline our approach to a solution. We mostly focus on generating a small, bare cosmological constant related to the cosmic angular momentum constant $J_0$. We defer analysis of quantum fluctuations of the vacuum to a later time, noting that developing a supersymmetric version of $TW$ gravity is an obvious avenue to consider. Alternatively, there are many non-supersymmetric ideas that may bear fruit with a merger of $TW$ gravity~\cite{Kachru:1998yy,Kachru:1998pg,Kachru:1998hd,Brandenberger:2002sk,Brandenberger:2004ki,Pejhan:2018afk,Alexander:2018tyf}.

 Since projective connections are central in the TW gravity approach, we give a brief projective geometry primer in section \ref{s:ProjectivePrimer}. The salient ingredients needed to discuss cosmology in the framework of TW gravity are laid out.   These ingredients are then used in  section~\ref{s:KGB} to construct  the  pure (i.e no explicit Einstein-Hilbert action) Gauss-Bonnet $TW$ action using the dynamics discussed in~\cite{Brensinger:2017gtb}. By starting with the projective Gauss-Bonnet action,  and using a natural    decomposition  of the diffeomorphism field,
\begin{align}\label{e:Decomp}
 \cD_{ab} = D_{ab} + \Lambda \frac{4}{\rd(\rd-1)}  g_{ab}, 
\end{align} 
 we generate an Einstein-Hilbert action with bare cosmological constant $\Lambda$, and an interaction term that couples \(D_{ab}\) to the metric.  $D_{ab}$ will further be decomposed into a traceless and trace term. We show that the scalar field proportional to  the trace, \( D=D_{ab} g^{ab}\), yields a theory which is free from ghosts and tachyons.
We  derive the field equations and stress-energy tensor for $D_{ab}$ from the $TW$ action.

Section~\ref{s:Solutions} contains our result that the vacuum solutions require the bare cosmological constant be related to the parameter $J_0$ of the $TW$ action. We dub the parameter $J_0$ as the cosmic angular momentum constant, as its relation to the cosmological constant is given by
\begin{align}
        J_0 = \frac{3 c^3}{32 \pi G \Lambda} \sim 10^{86}~\text{J}\cdot\text{s}~~~.
\end{align}
In section~\ref{s:J}, using various cosmic rotation measurements, we estimate a range of values for the upper bound of the angular momentum of the observable Universe $J_{\text{Obs}}$:
\begin{align}
        J_{\text{Obs}} \lesssim10^{79}\text{J}\cdot\text{s} ~ - 10^{91}\text{J}\cdot\text{s}~~~.
\end{align}
Clearly, $J_0$ fits within this range and can be thought of as a plausible cosmic angular momentum scale.

In the last section it is shown how we couple the projective connection to fermions and arrive at the Dirac equation in the presence of the diffeomorphism field. We find that one has the usual gravitational interaction arising from the spin connection plus an axial scalar coupling to the trace of the diffeomorphism field. This has implications for both dark matter and as a portal from fermions to dark matter.

 Our conventions and dimensions of the various constants and fields are summarized in appendix~\ref{a:Conventions} and the beginning of appendix~\ref{a:Cosmo}. Appendix~\ref{a:Cosmo} gives a general review of general relativity and cosmology, including recent results from measurements of relevant cosmological parameters.  The rest of the appendices explicitly show our derivation of the \(\rd\) dimensional TW action from the \((\rd +1)\)-dimensional action, the extraction of Einstein-Hilbert gravity sourced by $D_{ab}$ and $\Lambda$ through the decomposition in  Eq.~(\ref{e:Decomp}), derivations of the equations of motion and stress tensor for $TW$ gravity, and a proof that the stress tensor is divergence free.
\section{The Cosmological Constant Problem}\label{s:CCProblem}
Here we review the cosmological constant problem and our proposed method to investigate solutions via $TW$ gravity\cite{Brensinger:2017gtb}. A more complete review of the cosmological constant problem is given in~\cite{Weinberg:1988cp}. In appendix~\ref{a:Cosmo}, we summarize general relativity and cosmology in a Friedmann-Lemaitre-Robertson-Walker background, describing the calculation of the cosmological constant using current data. The simplest description of the cosmological constant problem comes from dimensional analysis of the cosmological constant. As the cosmological constant has units of curvature, or inverse area, its ``natural'' value constructed from fundamental constants would be one over the Planck length squared
\begin{align}\label{e:Lambdalp}
        \Lambda \approx l_{Pl}^{-2} = &  3.829 \times 10^{69}~\text{m}^{-2}
\end{align}
where the Planck length is $l_{Pl} = \sqrt{\hbar G/c^3} = 1.616 \times 10^{-35}~\text{m}$. This \emph{natural value} is famously roughly 120 orders of magnitude larger than the measured value~\footnote{The discrepancy between Eqs.~(\ref{e:Lambdalp}) and~(\ref{e:LambdaValue}) is more precisely 121 orders of magnitude. Taking instead $\Lambda$ to be proportional to the reduced Planck mass squared $\Lambda \sim M_{Pl}^2 c^3/\hbar \sim 10^{68}~\text{m}^{-2}$ where the reduced Planck mass is $M_{Pl} = \sqrt{\hbar c/(8 \pi G)} \approx 4.341 \times 10^{-9}~\text{kg}$ results in a 120 order of magnitude discrepancy from Eq.~(\ref{e:LambdaValue}).}
\begin{align}\label{e:LambdaValue}
        \Lambda \approx 1.2 \times 10^{-52} \text{m}^{-2}~~~.
\end{align}
This simple derivation illustrates at least a partial possible solution: find an appropriate angular momentum parameter, other than $\hbar$, that predicts Eq.~(\ref{e:LambdaValue}). This alternative angular momentum parameter would have to be enormously larger than $\hbar$ and in this paper, we argue that such an enormous angular momentum parameter arises naturally from $TW$ gravity.

Simply choosing an appropriately sized angular momentum parameter is only part of the solution, as summarized nicely by Weinberg~\cite{Weinberg:1988cp} where an expected value of the cosmological constant is demonstrated to arise from particle physics. In quantum field theory, the mass density of the vacuum in curved space-time is non-zero $\vevrho$ and gives rise to an energy momentum tensor for the vacuum given by
\begin{align}
        \Theta^{ab}_{\text{vac}} =&\vevrho c^2  g^{ab}
\end{align}
So even in a vacuum, the right hand side of Einstein's equations will not be zero and Einstein's equations would be instead
\begin{align}\label{e:EECC}
         R^{ab} - \frac{1}{2} g^{ab} R + g^{ab}\Lambda  = -\frac{8 \pi G}{c^2} \vevrho g^{ab}~~~.
\end{align}
Rearranging, we see that $\vevrho$ adds a contribution to the cosmological constant, forming an effective cosmological constant $\Lambda_{\text{eff}}$
\begin{align}
        \label{e:EELambdaEffective}
        &R^{ab} - \frac{1}{2} g^{ab} R + g^{ab}\Lambda_{\text{eff}}  = 0\\
        \label{e:LambdaEffective}
        &\Lambda_{\text{eff}}  = \Lambda +  \frac{8 \pi G}{c^2}  \vevrho 
\end{align}
The original constant $\Lambda$ is sometimes referred to as the bare cosmological constant. The cosmological constant problem is that the vacuum density $\vevrho$ is calculated to be much larger than the measured value of $\Lambda_{\text{eff}}$, that we previously called $\Lambda$ in Eq.~(\ref{e:LambdaValue})
\begin{align}\label{e:LambdaefValue}
        \Lambda_{\text{eff}} \approx 1.2 \times 10^{-52} \text{m}^{-2}.
\end{align}
The vacuum density can be estimated as the following integral with quantum gravity scale momentum cutoff of $p = M_{Pl} c$ where $M_{Pl}$ is the reduced Planck mass $M_{Pl} = \sqrt{\hbar c/(8\pi G)} \approx  4.341 \times 10^{-9}~\text{kg} $
\begin{align}
         \frac{8 \pi G}{c^2} \vevrho  =& \frac{8 \pi G}{c^2}  \frac{4 \pi }{(2 \pi \hbar)^3 c}\int_0^{M_{Pl} c } dpp^2 \frac{1}{2} \sqrt{ p^2 + m^2 c^2} \cr
         \approx & M_{Pl}^2 \frac{c^2}{16 \pi^2 \hbar^2} = 9.6 \times 10^{65} \text{m}^{-2}
\end{align}
for $m << M_{Pl}$.
These vacuum contributions are $118$ order of magnitude larger than the measured effective cosmological constant $\Lambda_{\text{eff}}$, thus it is considered \emph{unnatural} to choose the bare parameter $\Lambda$ in Eq.~(\ref{e:LambdaEffective}) on the scale of the contribution from $< \rho>$ but with a discrepancy that is fine tuned to be 118 orders of magnitude smaller. 

We separate the cosmological constant problem into the following two parts, focusing in this paper on the first part:
\begin{enumerate}
        \item Use projective geometry to provide a mechanism that produces a small, bare cosmological constant.
        \item Uncover ``beyond the Standard Model physics" that cancels all vacuum contributions from quantum field theory.
\end{enumerate}
By using $TW$ gravity to examine the first problem, we are exploiting a symmetry in Einstein's equations associated with geodesics and using a gauge principle to dictate the form of the Lagrangian.  Furthermore,  TW gravity makes contact with structures found in 2D quantum gravity\cite{Alekseev:1988ce,Alekseev:1988vx,Rai:1989js,Rodgers:2003an} through the  coadjoint orbits of the Virasoro algebra\cite{Witten:1987ty}.  The cosmological constant arises as a natural decomposition of the  associated gauge field, \(  \cD_{ab} \),  which has been dubbed the diffeomorphism field in the physics literature and is known as the projective Schouten tensor by differential geometers.   We do not address the second problem fully. However, we do  discuss how projective geometry interacts with fermions and find that an axial scalar coupling to all fermions can serve as a portal for dark matter. We note here that an obvious avenue to address the second problem would be to use supersymmetry which automatically has a vacuum energy of zero.  However, in  a Universe such as ours, where supersymmetry is clearly broken, it is not known how to maintain this zero of vacuum energy below supersymmetry breaking scales. Nonetheless, we wish to investigate a supersymmetric version of $TW$ gravity in the future to address problem two above.

\section{ Projective Geometry Primer }\label{s:ProjectivePrimer}
In string theory, the coadjoint orbits of the Virasoro algebra and affine Lie algebras  gave rise to geometric actions that are identified as the Polyakov 2D quantum gravity action and the Wess-Zumino-Witten model\cite{Alekseev:1988ce,Alekseev:1988vx,Rai:1989js,Witten:1987ty}. Associated with the Polyakov action is a background field, \( \cD_{ab} \), and with the affine Lie algebra  another background field \( \mathcal A_{a} \). Although \( \cA_a \) can easily be related to a Yang-Mills potential which has fundamental roots in the Lie algebra,  \( \cD_{ab} \) is often taken as a composite field of fundamental fields and an energy-momentum tensor that transforms anomalously under conformal transformations.  Because of this interpretation,  \( \cD_{ab} \) was historically external to gravity.  TW gravity was born out of theoretical investigations \cite{Branson:1996pe,Branson:1998bc,Rodgers:2003an,Rodgers:2006ep} that sought to put \( \cD_{ab} \) on the same footing as \( \cA_a \)  where it was also  fundamental and directly related to gravitation. With the interpretation of  \( \cD_{ab} \) as a projective connection in TW gravity,  its fundamental gravitational origins have been achieved \cite{Brensinger:2017gtb}. 

There are many excellent reviews and discussions of projective geometry\cite{Roberts,Eastwood,Crampin,OvsienkoValentin2005Pdgo} so this section will only give a pragmatic discussion on how one constructs the projective connection, the curvature tensors, the spin connection and how to build a metric that can be used to solder these constructs together to form an action.      

\subsection{The TW Projective Connection}\label{s:ProjectiveConnection}
Here we briefly describe the projective connection and explicitly show the construction of the TW covariant derivative operator. This will set us up to study cosmology in the context of projective geometry. 

Projective geometry arose from the question of connection ambiguities in geodesics on a manifold, say $\cal M$ \cite{Weyl,Thomas:1925a,Thomas:1925b,Whitehead:1931a,Cartan:1923zea,Cartan:1924yea}.  Since objects moving along geodesics is a principal way for physicists to infer the underlying metric, the question also has experimental relevance.  Two affine connections are said to be projectively related on $\cal M$, when there exists a one-form with components $ A_i$ such that 
\be
{\hat \G}^{i}_{\,\,jk}= \G^{i}_{\,\,jk}+ \d^i_k A_j+ \d^i_j A_k. \label{TransConnection}
\ee
Connections which are related in this way give rise to the same geodesics and are said to be projectively equivalent. 

Let's suppose that  $\cal M$ is a $\rd$-dimensional manifold. Projective geometry \cite{BaileyT.N.1994TSBf,Crampin,Roberts},  can then be cast as a gauge theory over \( \mathcal M \) giving rise to  $(\rd+1)$-dimensional manifold called the\emph{ Thomas Cone.}  It is equipped with a Thomas-Whitehead connection, \(   {\tilde \nabla}(\tilde \Gamma^{\alpha}_{\,\,\,\beta \gamma}) \) 
  \cite{Thomas:1926c}.   The extra dimension arises from adding a ``volume'' dimension with a new real coordinate $\l $ which takes values $0 < \l < \infty$. The coordinates on the $(\rd+1)$-dimensional Thomas cone are now denoted as \( x^\a= (x^0, \cdots , x^{\rd-1}, \l).\)  Throughout this paper we will use Greek indices to represent the full $(\rd+1)$  coordinates and Latin indices to represent the restriction to coordinates of \( \mathcal M \). On the Thomas Cone, there exists a preferred vector field, $\Upsilon$, which generates the  projective transformations through its Lie Derivative, where, for example,\be \mathfrak{L}_{\Upsilon} h=\ \Upsilon^\a \partial_\a h= \l \,\partial_\l h, \label{lambda}\ee   
 for  a function $h$.  This Lie derivative will vanish when $h$ is a projective invariant. There is also a preferred one-form $\o$ on the Thomas cone, which is related to $\Upsilon$ by the conditions that  $\o_\a \Upsilon^\a=1  $ and $\mathfrak{L}_\Upsilon \o_\rho =0$.      
From the volume form,  \[vol(\l) =  f(\ell) \epsilon_{a_1 \cdots a_n}dx^{a_1}\cdots dx^{a_n},\]  the relationship between  $\l$ and the volume   is established through a function $f(\ell)$ where  the  parameter $ \ell\equiv \frac{\l}{\l_0}$ is  dimensionless and   $\l_0$ is a constant.     
The projective connection and $\Upsilon$ are compatibly related by,
\be {\tilde \nabla}_\a \Upsilon^\b = \d_\a^\b .\ee
By explicitly writing  the pair, $\Upsilon^\alpha$ and $\omega_\alpha$ as,
\begin{align}
        \Upsilon^{\alpha} = (0,0,\dots,\lambda)~~~\text{and}~~~\omega_{\alpha} = (0,0,\dots, \lambda^{-1}),
\end{align}
the connection coefficients , ${\tilde \G}^{\b}_{\,\,\,\rho \,\a}$  may be written as   \cite{Roberts}: 
\begin{equation} 
{{\tilde\G}}^{\a}_{\,\,\b \g}= \begin{cases}
    {\tilde \G}^{\lambda}_{\,\,\,\lambda a}={\tilde \G}^{\lambda}_{\,\,\, a \lambda} = 0
    \\ {\tilde \G}^{\a}_{\,\,\,\,\lambda \lambda} = 0 \label{e:Gammatilde}\\ {\tilde \G}^{a}_{\,\,\,\,\lambda b}={\tilde \G}^{a}_{\,\,\,\,b \lambda} = \o_\lambda\,\d^a_b\\
{\tilde \G}^{a}_{\,\,\,\,b c} ={ \Pi}^{a}_{\,\,\,\,b c}\\
{\tilde \G}^{\lambda}_{\,\,\,\, a b} =  \Upsilon^\lambda \cD_{ a b}
 \end{cases} 
\end{equation} 
where the projective invariant connection \( \Pi^a_{\,\,\,\,bc} \) is defined
\begin{equation}{ \Pi}^{a}_{\,\,\,\,b c} ={ \G}^{a}_{\,\,\,\,b c} - \frac{1}{d+1}(\delta^{a}_{\,\,\,\, c} { \G}^{d}_{\,\,\,\,b d} + \delta^{a}_{\,\,\,\, b} { \G}^{d}_{\,\,\,\,c d})   \end{equation}
  with ${\G}^{b}_{\,\,\,c \,a}$ the connection coefficients  on the spacetime $\mathcal M$.  The projective connection \( \cD_{ab} \)  is independent of the  coordinate $\lambda$ and transforms as a rank two tensor with  an additional inhomogeneous term related to the Jacobian of the transformation on  $\cal M$. To be precise, consider a coordinate transformation on the Thomas Cone given by   
\begin{equation}
 p^\a=( p^0, p^1, \cdots p^{\rd-1},\l) \rightarrow q^\a=(q^0(p), q^1(p), \cdots q^{\rd-1}(p),\l   J(q,p)^{\frac{-1}{d+1}})),
\label{CoordinateTransformation} \end{equation}
 where $J(q,p)= |\frac{\partial q^i}{\partial p^j} |$ corresponds to the determinant of the Jacobian of the transformation of the  coordinates on $\cal M$.  Then, in order for ${\tilde \nabla}_\a $ to transform as an affine connection,  ${ \cD}_{a b}$ must transform as
\be
{\cD}_{a b}' = \frac{\partial p^c}{\partial q^a} \frac{\partial p^d}{\partial q^b}{\cD}_{c d} +  \frac{\partial p^l}{\partial q^c}( \frac{\partial^2 q^c}{\partial p^l \partial p^d} \frac{\partial^2 p^d}{\partial q^a \partial q^b }) + \frac{\partial q^m}{\partial p^{n}} \frac{\partial^3 p^n}{\partial q^m \partial q^a \partial q^b},
\label{Difftransformation} \ee
when $( p^0, p^1, \cdots p^{\rd-1})\rightarrow (q^0, q^1, \cdots q^{\rd-1})  $ on $\cal M$.  From here, one can construct the projective curvature tensor that remains invariant  when a connection on $\cal M$ transforms as Eq.~(\ref{TransConnection}) and is covariant under the coordinate transformation Eq.~(\ref{CoordinateTransformation}). In the physics literature, \( \cD_{ab} \) is called the diffeomorphism field and the projective Schouten tensor in  differential geometry.  Two connections, then,  are in the same projective equivalence class, $[ \G^a_{~~bc}] = [\hat \G^a_{~~bc}]$, when they have the same projective curvature tensor.  We  explicitly construct the projective curvature tensor in what follows.    
\subsection{Projective Curvature}\label{s:Curvature}
Using the explicit construction of projective connection coefficients, it is straightforward to compute the curvature invariants.  Explicitly, on a vector field \(\kappa^\a\) and co-vector \(\kappa_\a\) on the Thomas cone,  we define the projective curvature tensor ${K}^{\g}_{\,\,\,\rho\a \b  } $ in the usual way,
\be 
[{\tilde \nabla}_\a,{\tilde \nabla}_\b] \kappa^\g = \,{K}^{\g}_{\,\,\,\rho\a \b  } \kappa^\rho~~~\text{and}~~~[{\tilde \nabla}_\a,{\tilde \nabla}_\b] \kappa_\g =- \,{K}^{\rho}_{\,\,\,\g\a \b  } \kappa_\rho.
\ee
In terms of the connection coefficients, 
\begin{align}\label{e:PCurv}
        K^\mu{}_{\nu\alpha\b} \equiv \tilde{\G}^\mu{}_{\nu[\b,\a]} + \tilde{\G}^\rho{}_{\n[\b}\tilde{\G}^\mu{}_{\a]\rho}~~~.
\end{align}
Using Eq.~(\ref{e:Gammatilde}),  the only non-vanishing components of $K^\mu{}_{\nu\alpha\b}$ are
\begin{align}
        K^a{}_{bcd} = \mathcal{R}^a{}_{bcd} + \delta_{[c}{}^a \cD_{d]b}~,&~~~ \\
        K^\l{}_{cab} = \lambda \partial_{[a} \cD_{b]c} + \lambda \Pi^d{}_{c[b}\cD_{a]d}~.&~~~
        \label{e:KRiemannZero} \end{align}
Here
\begin{equation}
\mathcal{R}^a{}_{bcd}= \partial_c \Pi^{a}_{\,\,\,\,db}- \partial_d \Pi^{a}_{\,\,\,\,c b} + \Pi^{a}_{\,\,\,ce}\Pi^{e}_{\,\,\,db}-\Pi^{a}_{\,\,\,de}\Pi^{e}_{\,\,\,cb}
\end{equation}  is the curvature produced from the $\Pi^a_{\,\,\,\,bc}$ connections. Although $K^a{}_{bcd}$ and $K^\l{}_{cab}$ are tensors under general coordinate transformations on \( \mathcal M \), \(\mathcal{R}^a{}_{bcd} \) and \( \mathcal D_{ab}\) are not.  However, both \(\mathcal{R}^a{}_{bcd} \) and \( \mathcal D_{ab}\) are tensors under volume-preserving diffeomorphisms. Since  we are interested in metric compatible connections in what follows, we fix the coordinates in a constant volume gauge so that Eq.~(\ref{e:KRiemannZero}) becomes:  
\begin{align}
        K^a{}_{bcd} = {R}^a{}_{bcd} + \delta_{[c}{}^a \cD_{d]b}~,&~~~ \\
        K^\l{}_{cab} = \lambda \partial_{[a} \cD_{b]c} + \lambda \G^d{}_{c[b}\cD_{a]d}~.&~~~
        \label{e:KRiemann} \end{align} 
From here we calculate the only non-vanishing components of the projective  Ricci tensor
$K_{\alpha\beta}\equiv K^\rho{}_{\alpha\beta \rho}$ and the projective scalar curvature $K \equiv K_{\alpha\beta}G^{\alpha\beta}$ (the metric will be defined in a moment) to be
\begin{align}
        K_{ab} & = R_{ab} - (\rd-1) \cD_{ab}~~~,~~~K=G^{ab} ( R_{ab} - (\rd-1) \label{ProjectiveScalar} \cD_{ab}).
\end{align}
   In the above $R^a{}_{bcd}$ is the Riemann curvature tensor over the manifold $\mathcal{M}$, defined in terms of its connection coefficients, $\Gamma^{a}{}_{bc}$.  
It is important to note that we have not yet defined the   $\rd+1$ dimensional metric $G_{\a\b}$.  Its construction will be made explicit in the following section \ref{s:SpinConnection}.
. 
\subsection{Projective Metric and Spin Connection}\label{s:SpinConnection}

We now proceed to construct the metric, $ G_{\a \b}$, for the \((\rd+1)\)-dimensional manifold,  that was alluded to in Eq.~(\ref{ProjectiveScalar}). 
Let's assume for the moment that $\rd$ is even. The Dirac matrices are related to a metric $g_{ab}$ on the spacetime manifold, $\mathcal{M }$, by \be \{\g^a, \g^b\} = 2 g^{a b}.\ee 
   As stated above, 
 we will write the indices related to coordinates on $\mathcal{M}$ as $a,b = 0,\cdots ,\rd-1,$ where $\rd$ is the dimension of the manifold. We can define an extra gamma matrix, $\g_{\rd}$ (with index down) that is related to the volume parameter $\l$  via, 
\be
\gamma(\l)_{\rd} = \epsilon \, \frac{f(\ell)}{\rd !} \epsilon_{a_0 \cdots a_{\rd-1}}\g^{a_0}\cdots \g^{a_{\rd-1}},
\ee
with $\epsilon$ chosen  to be $1$ 
so that the new direction is space-like in the constructed metric.    Then an extended metric can be defined on the Thomas cone through
\be \{\g_\a, \g_\b\} = 2 G_{\a \b}~ \bf 1,\ee  
 where  $\a,\b = 0,\cdots ,\rd ,  $ \(\bf 1 \) is the fermion identity, and   \begin{align}
        G_{\a \b} =& \left( 
                        \begin{array}{cc}
                                g_{ab} & 0 \\
                                0 & -f(\ell)^2
                        \end{array}
        \right) \label{e:Gmetric} \\
        G^{\a \b} =& \left( 
                        \begin{array}{cc}
                                g^{ab} & 0 \\
                                0 & -f(\ell)^{-2}
                        \end{array}
        \right). \label{e:GmetricInverse}
\end{align}
For our purposes in four dimensions, we have chosen $\epsilon$ above so the chiral matrix, $\g_5 \equiv i \g^4.$   This will later guarantee that the spinor connection, \( i\, \O_\mu\) defined below is self-adjoint.  Metrics of this form have been used in the literature  to study other projective properties of Einstein manifolds, 
geodesics paths on Einstein spaces, higher spin fields and Bernstein-Gelfand-Gelfand complexes \cite{CapA.2014Emip,Gover:2014vxa,GoverA.Rod2012DEgE}. Although this Dirac matrices construction required $\rd$ to be even, this form of the metric \( G_{\a \b}\)  can be used in any dimension.  For us, the relationship with the volume and chirality becomes pronounced when we include fermions. In what follows, we will use $G_{\a \b}$ to contract with the  projective curvature for the interaction Lagrangian and the dynamical action for the diffeomorphism field. 

We proceed with the construction of the spin connection on the Thomas cone. $G_{\a \b}$ admits frame fields through, 
\be
G_{\m \n} = e_{\m}^{\,\,A} e_{\n}^{\,\,B} \eta_{A B}\,\,\,\, \text{and   } \eta_{A B} = g_{\m \n}E^{\m \,\,}_{\,\,A}  E^{\n \,\,}_{\,\,B},
\ee  
where the ``flat'' indices, \( A, B = 0 \ldots \rd\). Since the projective connection is incompatible with a metric, we define the spin connection for the projective connection and the frame fields through, \be
 {\tilde\o}_{ A B \a} =e_{\n}^{ \,C} ({ \partial}_{\a}E^{\n}_{\,\, B}+\tilde{\G}^\nu_{\,\,\alpha \beta } E^{\b}_{ B}) \,\eta_{A C}.  
  \ee 
This guarantees that
\be
{\tilde \nabla}_\m  E^{\b}_{\,\, B} \equiv \partial_\m E^{\b}_{\,\, B} +\tilde{\G}^\b_{\,\,\m \a } E^{\a}_{ B} - {\tilde\o}^{ A}_{\,\, \,B \m}E^{\b}_{\,\, A}=0.
\ee
\ 
For transparency, let us write the spin connection in terms of the four spacetime dimensions and the volume direction explicitly.  The flat directions  will be denoted by $a,b$ for the spacetime directions and the number $``4"$ for the flat volume direction.   Similarly, we will use $\m, \rho$ for the spacetime coordinates and reserve $``\l"$ for the volume direction on the Thomas cone.  With this we may write the projective spin connection as
  
\be
\tilde\o_{A B \m}= \begin{cases}
    \o_{a b \m},& A=a, B=b,\m= 0, \cdots, 3  \\ \frac{1}{\l} \eta_{a b} ,& A=a, B=b,\m=\l  \\ -\frac{1}{\l f(\ell)}e_{\m}^{\,\,c} \eta_{a c},& A=a, B=4,\m= 0, \cdots, 3    \\
\l f(\ell) {\mathcal D}_{\m \rho}E^{\rho}_{\,\,\, b}  ,& A=4, B=b,\m= 0, \cdots, 3\\ 0 & A=a, B=4,\m= \l \\ 0 & A=4, B=b,\m= \l \\  0 & A=4, B=4,\m=0, \cdots, 3 .
  
\end{cases} \label{spinconnection}
\ee

\section{The Diffeomorphism Field Action}\label{s:KGB}
Using the metric $G_{\a \b}$ from above, the determinant $G = \det(G_{\a\b})$ and its square root are
respectively,\begin{align}\label{e:Gdet}
        G = -g f(\ell)^2~~~\text{and}~~~\sqrt{|G|} = \sqrt{|g|}f(\ell),
\end{align}
where $g = \det{(g_{ab})}$ and $|g|$ is the absolute value of $g$.  From  the non-vanishing components of $K^{\alpha}_{~\beta\mu\nu}$, Eq.~(\ref{e:KRiemann}) and Eq.~(        \ref{e:KRiemann}) become,
\begin{align}        \label{e:KRicci}
        K_{ab} =& K^{\mu}{}_{ab\mu} =  R_{ab} -(\rd - 1) \cD_{ab} \\
        \label{e:K}
        K\equiv & G^{\alpha\beta}K_{\alpha\beta} = R - (\rd - 1)\cD \\
        \label{e:RD}
                R =& g^{ab}R_{ab}~~~,~~~ \cD =  g^{ab}\cD_{ab}~~~.
\end{align}
In performing these calculations, it is important to keep in mind that the symmetry properties of $K_{\alpha\beta\mu\nu} = G_{\alpha\rho}K^{\rho}{}_{\beta\mu\nu}$ and $R_{abcd} = g_{am}R^{m}{}_{bcd}$ are not the same. For instance $R_{abmn} = - R_{bamn}$ but $K_{\alpha\beta\mu\nu} \ne -K_{\beta\alpha\mu\nu}$. This is due to the connection $\tilde{\Gamma}^{\alpha}{}_{\mu\nu}$ being incompatible with $G_{\mu\nu}$ although $\Gamma^{a}_{mn}$ is compatible with $g_{mn}$. The astute reader will realize that $K^c{}_{\lambda ab} = 0$ while $K^\lambda{}_{cab} \ne 0$. The complete symmetries of $K_{\alpha\beta\mu\nu}$ and $R_{abcd}$ are
\begin{align}
        \label{e:KRiemannSymmetries}
        K_{\alpha\beta\mu\nu} =& - K_{\alpha\beta\nu\mu} \\
        R_{abcd} = - R_{bacd} =& R_{badc} = R_{cdab}
\end{align}
The rank three tensor, $K_{\alpha\beta\gamma}$, is called the projective Cotton-York tensor and is defined as
\begin{align}
        K_{\alpha\beta\gamma} \equiv K^{\rho}{}_{\alpha\beta\gamma}\omega_\rho = \lambda^{-1} K^\lambda{}_{\alpha\beta\gamma}. 
\end{align}
The only non-vanishing components of $K_{\alpha\beta\gamma}$ are
\begin{align}\label{e:K3}
        K_{bac}= \nabla_a \cD_{b c}-\nabla_c \cD_{b a}~~~.
\end{align}
This satisfies the Bianchi identity
\begin{align}\label{e:KBianchi}
       K_{[acm]} = & K_{acm} + K_{cma} + K_{mac} = 0~~~.
\end{align}
The action of projective Gauss-Bonnet with coupling constant $J_0$, which we refer to as $TW$ gravity, is given by
\begin{align}\label{e:STWdplus1}
        S_{TW} =&  - \tfrac{J_0 c}{2}\int d\ell\; d^{\rd} x\sqrt{|G|} \left[K^2 - 4 K_{\alpha\beta}K^{\alpha\beta} + K_{\alpha\beta\mu\nu}K^{\alpha\beta\mu\nu}   \right]
\end{align}
Interestingly enough only the measure depends on the parameter $\ell(\l)$. Therefore we can reduce  the  $(\rd +1)$-dimensional action, above to $\rd$-dimensions by integrating out the $\ell$-dependence. As shown in appendix~\ref{a:Ldetails}, all $\ell$-integrations take the form of one of the two integrals below, the first of which we normalize to one, the other we define through a new constant $\alpha_0$:

\begin{align}
        \int_{\ell_i}^{\ell_f} d\ell f(\ell) = 1~~~,~~~\alpha_0=& \lambda_0^2 \int_{\ell_i}^{\ell_f} d\ell \ell^2 f(\ell)^3~~~.
\end{align}
Once we choose $f(\ell)$, and properly normalize to satisfy the first integral, this will fix $\alpha_0$ in terms of $\ell_i$ and $\ell_f$. As shown explicitly in appendix~\ref{a:Ldetails}, using the above expansions of the projective curvature $K^{\alpha}_{~\beta\mu\nu}$ and the metric $G_{\mu\nu}$, the $TW$ action can be written as
\begin{align}\label{e:STWa}
    S_{TW} =&  \int d^\rd x\sqrt{|g|} \mathcal{L}_{\cD}  + S_{GB}~~~, \\
    \label{e:LDa}
   \mathcal{L}_{\cD} = & \tfrac{J_0 c}{2}\left[\alpha_0  K_{bmn}K^{bmn} -  \cD_{ab}\tcD_*^{ab} \right] ~~~,  \\
    \label{e:SGBa}
    S_{GB} = & -\tfrac{J_0c}{2} \int d^\rd x \sqrt{|g|} \left( R^2 - 4 R_{ab}R^{ab} + R_{abmn}R^{abmn} \right)~~~
\end{align}
where
\begin{align}
\label{e:tcDstar}
        \tcD_*^{ab} =& (\rd -1 ) g^{ab} \tcD - 2 (2 \rd- 3)\tcD^{ab}\\
        \label{e:tcDab}
        \tcD_{ab} =& (\rd - 1)\cD_{ab} -2 R_{ab} = -K_{ab} - R_{ab}\\
        \label{e:tcD}
        \tcD = g^{ab}\tcD_{ab} =& (\rd - 1)\cD - 2 R = -K - R
\end{align}

Generally, we define the star $(*)$ operation on an arbitrary rank-two tensor as
\begin{align}\label{e:Star}
        T_*^{ab} =& M^{abmn} T_{mn} = (\rd - 1) g^{ab} T - 2 (2 \rd - 3) T^{ab} 
\end{align}
where $T = g^{mn}T_{mn}$ and the $M^{abmn}$ tensor is
\begin{align}\label{e:Mtensor}
        M^{abmn} =& (\rd -1) g^{ab}g^{mn} - 2 (2 \rd - 3) g^{am}g^{bn}~~~.
\end{align}
This tensor is symmetric under any permutation involving all four of its indices:
\begin{align}\label{e:MSymmetries}
        M^{abmn} =& M^{banm} = M^{mnab} = M^{nmba}~~~.
\end{align}
At this point it is important to observe that the $S_{TW}$  action is a function of three dynamical variables, viz \( g_{ab}, \G^{a}_{~~bc},\) and \(\cD_{ab} \). The field equations for  the metric and connection can be examined  independently \cite{Borunda:2008kf,BasteroGil:2009cn,Bernal:2016lhq,palatini} in the context of Gauss-Bonnet, and because we are in four-dimensions, metric compatibility is still a solution to the   field equations.   Also, we have already  mentioned in section \ref{s:ProjectivePrimer} that when one of the members of a projective equivalence class $[\G^a_{~bc}]$ is a metric compatible connection, the projective Schouten tensor collapses to a constant times the metric \cite{Gover:2014vxa,GoverA.Rod2012DEgE}.   \( \cD_{ab} \) is that projective  Schouten tensor when it is not dynamical. We exploit this observation when we promote  \( \cD_{ab} \) to a  dynamical field by separating out a part that vanishes in the projective Cotton-York tensor, Eq.~(\ref{e:K3}), from the non-trivial dynamical degrees of freedom.  It is natural, therefore, to  write the diffeomorphism field as,  
\begin{align}\label{e:Shift}
        \cD_{ab} = &D_{ab} + \frac{4}{\rd (\rd - 1)} \Lambda g_{ab} \cr
               = &\left[W_{ab} + \frac{mc}{J_0 \rd}g_{ab} \phi \right] + \frac{4}{\rd (\rd - 1)} \Lambda g_{ab} ~~~
\end{align} when we assume \( \nabla_a g_{bc} =0\). 
Here
 $W_{ab}$ is traceless
\begin{align}
        g^{ab}W_{ab} = 0
\end{align}
and $\Lambda$ is the bare cosmological constant. The parameter $m$ is the mass of the scalar field $\phi$, arising from the trace of  $D_{ab}$.  This decomposition will naturally produce  an Einstein-Hilbert action with cosmological constant and an accompanying interaction for the dynamical degrees of freedom, by starting with only the projective Gauss-Bonnet Eq.~(\ref{e:LDa}).  In this paper, we will analyze the $TW$ action in terms of $D_{ab}$ and use the $\phi$ decomposition to show an absence of ghost and tachyon pathologies. The rank three tensor $K_{abc}$ is the same whether written in terms of $\cD_{ab}$ or $D_{ab}$, owing to the covariant derivative. The decomposition, Eq.~(\ref{e:Shift}), does modify  $\mathcal{L}_{\cD}$ slightly by producing  terms involving $\Lambda$  in the product $\cD_{ab} \tcD_*^{ab}$. The details of this are found in appendix~\ref{a:Ldetails}.  The result is
\begin{align}\label{e:STW}
    S_{TW} =& \frac{1}{2\kappa} \int d^\rd x\sqrt{|g|} \left( R - 2 \Lambda\right)+ \int d^\rd x\sqrt{|g|} \mathcal{L}_D + S_{GB}~~~, \\
    \label{e:LD}
    \mathcal{L}_{D} 
    = & \tfrac{J_0 c}{2}\left[\alpha_0 K_{bmn}K^{bmn} - D_{ab}\tD_*^{ab} - y(\rd) \Lambda D \right]~~~,  
\end{align}
where $\kappa$ is the $\rd$-dimensional gravitational coupling related to the bare cosmological constant as
\begin{align}
        \label{e:const}
        \Lambda =& \frac{\rd -1}{J_0 c y(\rd) \kappa} ~~\text{and}\\
            \label{e:yd}
    y(\rd) =& 8 (\rd -2)(\rd-3)/\rd.
\end{align}
The new tensors appearing above are
\begin{align}
\label{e:tDstar}
        \tD_*^{ab} =& (\rd -1 ) g^{ab} \tD - 2 (2 \rd- 3)\tD^{ab}\\
        \label{e:tDab}
        \tD_{ab} =& (\rd - 1)D_{ab} -2 R_{ab} = -K_{ab} - R_{ab}\ - \tfrac{4}{\rd}\Lambda g_{ab}\\
        \label{e:tD}
        \tD = g^{ab}\tD_{ab} =& (\rd - 1)D - 2 R = -K - R - 4 \Lambda
\end{align}
For $\rd=4$ dimensions, $\kappa$ becomes the Newton constant, with appropriate factors of $\pi$ and $c$
\begin{align}\label{e:kappasquared}
        \kappa = 8 \pi G/c^4 ~~~\text{for}~~~\rd =4.
\end{align}

\subsection{Tachyons and Ghosts}
It is desirable that the consideration of the TW gravity action Eq.~ (\ref{e:STW}) not introduce any potential pathologies like ghost and tachyonic fields at the classical level. To determine the potential for such problematic fields, we analyze the dynamical trace degrees of freedom from the decomposition in Eq.~ (\ref{e:Shift}) in a Minkowski background. For the sake of this analysis, we will take the dimensionally extended metric Eq.~(\ref{e:Gmetric}) to be
\begin{align}
\label{e:TGmetric}
G_{\alpha \beta}=& \left( 
                        \begin{array}{cc}
                                \eta_{ab} & 0 \\
                                0 & A f(\ell)^2
                        \end{array}
        \right)~~~
\end{align}
where $A$ will ultimately be $\pm 1$. Ghosts and tachyons arise from the kinetic and potential terms so here we will only concern ourselves with terms quadratic in the fields. The relevant piece from the TW gravity Lagrangian is
\bea
\label{e:D_quad_decomp}
\mathcal{L}_{D^2}=\frac{J_0 c}{2}\left(-\alpha_0 A K_{bmn}K^{bmn}-D_{ab}\tilde{D}^{ab}_*\right).
\eea
Applying the decomposition $D_{ab}=\frac{mc}{J_0 d}\eta_{ab}\phi$ to this piece, we find
\begin{align}
\label{e:phi_decomp}
\mathcal{L}_{\phi^2}&=\frac{m^2 c^3}{2 J_0 \rd^2}\left[-\frac{\alpha_0}{2} A(\rd-1)\partial_{\mu} \phi \partial^{\mu}\phi-\rd(\rd-1)(\rd-2)(d-3)\phi^2\right]\\
&=\frac{m^2 c^3}{2 J_0 \rd^2}\left[-\frac{\alpha_0}{2} A(\rd-1)\left(\dot{\phi}^2 - (\nabla \phi)^2\right)-\rd(\rd-1)(\rd-2)(\rd-3)\phi^2\right].
\end{align}
where we have separated temporal and spatial components on the last line. To clearly determine the potential for ghost/tachyon fields, we need the corresponding Hamiltonian. Defining the conjugate momentum as
\bea
\pi \equiv \frac{\partial \mathcal{L}_{\phi^2}}{\partial \dot{\phi}}=-\frac{\alpha_0 A (\rd-1)m^2 c^3}{2J_0 \rd^2}\dot{\phi},
\eea
we can write the Hamiltonian as
\begin{align}
\mathcal{H}_{\phi^2}&=-\frac{\alpha_0 A (\rd-1)m^2 c^3}{2J_0 \rd^2}\dot{\phi}^2-\mathcal{L}_{\phi^2}\\
&=\frac{m^2 c^3}{2J_0 \rd^2}\left[-\frac{\alpha_0}{2} A (\rd-1)\left(\dot{\phi}^2+(\nabla \phi)^2 \right)+\rd(\rd-1)(\rd-2)(\rd-3)\phi^2 \right].
\end{align}
Since a negative kinetic term results in a ghost field, and a negative sign in the mass term results in a tachyonic field, we see that setting $A=-1$ results in a non-pathological classical field while setting $A=+1$ results in a ghost field, the field being non-tachyonic in either case. We expect this analysis to hold, at least perturbatively, for other metrics.  We note that in $\rd=4$, $\phi$ becomes a massive field suggesting a short range gravitational wave.  The phenomenology of these waves, along with the traceless components, will be part of a future study.

\subsection{Equations of Motion}
Here we present the equations of motion, detailed derivations are given in appendices~\ref{a:EQMDerivation} and~\ref{a:StressEnergyDerivation}. Variation of the action $S_{TW}$ with respect to the diffeomorphism field $D_{ab}$ yields its equations of motion
\begin{align}\label{e:EOM}
        \alpha_0 \nabla_n K^{(ab)n}  =& -K_*^{ab} ~~~.
\end{align}
where Eqs. (\ref{e:tDab}), (\ref{e:Star}), and (\ref{e:yd}) can be used to show that the source can be written as
\begin{align}
 K_*^{ab} =&  R_*^{ab} - (\rd - 1) D_*^{ab}  -\tfrac{1}{2} y(\rd)\Lambda g^{ab}  \cr
          =&-\tfrac{1}{2} [ \tD_*^{ab} + (\rd - 1) D_*^{ab} + y(\rd)\Lambda g^{ab} ]~~.
\end{align} 
Einstein's equations for the diffeomorphism field coupled to $g_{ab}$ are
\begin{align}\label{e:EE}
        R^{ab} - \frac{1}{2} g^{ab} R + g^{ab}\Lambda = - \kappa ~ \Theta^{ab}~~~.
\end{align}
In terms of $V^{cab} $, \begin{align}
        V^{cab} =&\alpha_0 K^{(ab)m}?D_m^c? + \tfrac{1}{2} g^{c(a} \nabla_m D_*^{b)m} - \nabla^c D_*^{ab}
\end{align}
the stress-energy tensor can be expressed as
\begin{align}
\begin{split}
        \Theta^{ab} =& J_0 c \left(\nabla_c V^{cab} -\nabla_c V^{(ab)c}\right)  + \tfrac{\alpha_0 J_0 c}{2} \left[?K^(a_mn? ?K^b)mn? + 2 ?K^cm(a? ?K_cm^b)? \right] \cr 
&- \tfrac{J_0 c}{2} ?D_c^(a? \tD_*^{b)c} -\tfrac{J_0 c}{2} D_*^{c(a} \tD?{}_c^b)? - J_0 c y(\rd)\Lambda D^{ab} -  g^{ab} \mathcal{L}_D + \Theta_{(GB)}^{ab}~~~
\end{split} \label{e:StressEnergy}
\end{align}
with $\mathcal{L}_{D}$  the Lagrangian density, Eq.~(\ref{e:LD}).  $\Theta_{(GB)}^{ab}$ is the contribution from $S_{GB}$ which in $\rd \le 4$ does not contribute to the classical field equations\cite{Lanczos:1938sf}.

\section{Vacuum Solution and Angular Momentum of the Universe}\label{s:Solutions}
\subsection{Vacuum Solution of the Equations of Motion}
In this section we demonstrate that for $\rd = 4$ the trivial solution $D_{ab}=0$ is consistent with both Eqs.(\ref{e:EOM}) and~Eq.~(\ref{e:EE}) and reduces these equations to the vacuum Einstein Equations sourced by a cosmological constant.  Recalling the decomposition in Eq.~(\ref{e:Shift}) and then setting $D_{ab} =0$ in the action in Eq.~(\ref{e:STW}) is tantamount to setting $D_{ab}$ proportional to the metric in the original action in Eq.~(\ref{e:STWdplus1}).    Setting $D_{ab}=0$  results  in the following simplifications
\begin{align}
 K^{ab} = R^{ab} - \tfrac{4}{\rd}\Lambda g_{ab}~~~,~~~K = R- 4\Lambda~~~,~~~K_{abc} = 0.
\end{align}
For $\rd \le 4$, the stress energy tensor, Eq.~(\ref{e:StressEnergy}), vanishes under these conditions (as $\Theta_{(GB)}^{ab} =0$), and Eq.~(\ref{e:EE}) reduces to the pure cosmological constant sourced vacuum Einstein Equations
\begin{align}\label{e:EECCVacSoln}
         R^{ab} - \frac{1}{2} g^{ab} R + g^{ab}\Lambda  = 0.
\end{align}
The $D_{ab}$ equation of motion Eq.~(\ref{e:EOM}) reduces to 
\begin{align}\label{EOMCC}
        R^{ab} - \frac{1}{2} \frac{\rd -1}{2\rd -3}g^{ab} R + g^{ab}\frac{y(\rd)}{4 (2 \rd  -3)}\Lambda  = 0,
\end{align}
where to cast the term involving the cosmological constant in this form, we have used Eq.~(\ref{e:const}). 
If we contract Eqs.~(\ref{e:EECCVacSoln}) and~(\ref{EOMCC}) with the metric, we arrive at the following constraints, respectively
\begin{align}\label{e:EEConstraints}
(\rd-2) R = 2 \rd \Lambda,\quad R(\rd-3)(\rd-2)=4 (\rd-3)(\rd-2)\Lambda.
\end{align}
The first equation in Eq.~(\ref{e:EEConstraints}) is the usual condition for an Einstein manifold with $R_{ab}=\frac{2}{\rd-2}\Lambda g_{ab}$ and the second equation is trivial for  $\rd=2$ and $\rd=3$ and leads to $R_{ab}=\frac{4}{\rd}\Lambda g_{ab}=\Lambda g_{ab}$ for $\rd=4$.  In $\rd=2$ we can readily see that the first equation implies $\Lambda=0$ and in $\rd=3$ we have that $R=6\Lambda$.
For $\rd > 4$, $\Theta^{ab}_{(GB)} \ne 0$ so the analysis presented in this section would not apply.
We will focus on $\rd=4$ in which we find a consistent solution to Eq.~(\ref{e:EEConstraints}) where $R_{ab}=\Lambda g_{ab}$ or equivalently $R=4\Lambda$. Then  Eq.~(\ref{e:const}) for the cosmological constant becomes
\begin{align}\label{e:Lambda4}
        \Lambda = &\frac{3}{4 J_0 c \kappa}~~~,~~~\text{for}~\rd = 4.
\end{align}
 Using the value for the Cosmological constant calculated from an average of the \emph{Planck} data and the \emph{Riess} collaboration, $\Lambda \approx 1.2 \times 10^{-52}~\text{m}^{-2}$ as shown in appendix~\ref{a:Cosmo}, we solve Eq.~(\ref{e:Lambda4}) for $J_0$:
\begin{align}\label{e:J0Lambda}
        J_0\approx 1.0 \times 10^{86} \text{J}\cdot\text{s}~~~\text{for}~~~\Lambda \approx 1.2 \times 10^{-52}~\text{m}^{-2}~~~.
\end{align}
Comparing with astronomical data as explained in the next section, the above value for the angular momentum parameter $J_0$ lies within the range for the upper limit of angular momentum of the observable Universe $J_{\text{Obs}}$ calculated from various measurements of cosmic rotation
\begin{align}
J_{\text{Obs}}\lesssim  10^{79}~\text{J}\cdot\text{s}  - 10^{91}~\text{J}\cdot\text{s}~~~.
\end{align}
We therefore now refer to $J_0$ as the cosmic angular momentum constant.

\subsection{Calculation of Expected Angular Momentum of the Observable Universe from Astronomical Data}\label{s:J}
In this section we briefly review the astronomical data suggesting a global rotation of the observable universe and from this data calculate a range of upper bounds to the angular momentum of the universe.
The Universe having patches of angular momentum that sum to zero is consistent with the cosmological principle. 
 In fact, taking the cosmic angular momentum constant $J_0$ as a fundamental constant would set the natural scale over patches where the Universe could have net angular momentum. At present there does appear to be some evidence for the rotation of the universe 
on large scales,
though we caution that this is somewhat controversial, as global rotation is difficult to measure and seems to be highly model dependent.  

Observational evidence of angular momentum of the present day universe on large scales has been seen in the parity violation of the angular momentum of spiral galaxies with a preferred axis \cite{Longo:2011}.
Models of global rotation using input from observations \cite{Li:1997du,Godlowski:2003} have been in agreement on the order of magnitude of the current angular rotation of the universe of $\omega\sim 10^{-13}~\text{rad}/\text{yr}$.
Another clear indication of rotation would appear in CMB data as anisotropy with a preferred axis.
New Planck data has found anisotropies at large angular scales at about the 2-3 $\sigma$ level that could be physically significant, see for example \cite{Schwarz:2015cma}.
A theoretical model using CMBA data constrained the rotation of the early universe to be $\omega\sim 10^{-9}~\text{rad}/\text{yr}$ \cite{Su2009}. 
A more conservative estimate using tighter constraints from both temperature and polarization data from Planck on Bianchi models of rotation \cite{Saadeh:2016sak} conclude that $\omega/H_0<10^{-11}$, which using the average value $H_0$ as in Eq.~(\ref{e:OmegaH}) and inserting a factor of $2\pi$ to convert to rad/yr yields $\omega \lesssim 10^{-21}~\text{rad}/\text{yr}$.  It should be noted that these values correspond to the rotation of the universe at the surface of last scattering and not the current value, which would be significantly lower.   It is possible, however, that only shear rotation can affect the CMB data and that global rotation may not influence CMB data.

Given these several pieces of evidence for rotation on cosmic scales, we present a simple order of magnitude estimation that demonstrates the cosmic angular momentum constant $J_0$ associated with the measured value of cosmological constant as in Eq.~(\ref{e:Lambda4}) is within the range of plausible angular momentum of the observable Universe.  If we approximate the observable Universe as a homogeneous rotating sphere of radius  $R_{\text{Obs}}=46.5\times 10^9~\text{lyr}$ and use the current estimate for mass density of the universe to be $\rho=10^{-26}\text{kg}/\text{m}^3$ we can calculate the total mass of the observable Universe $M_{\text{Obs}}=\frac{4}{3}\pi R_{\text{Obs}}^3\rho$ and moment of inertia $I_{\text{Obs}}=\frac{2}{5}M_{\text{Obs}} R_{\text{Obs}}^2$.  The total angular momentum then depends on the estimate of angular rotation $\omega$ as $J_{\text{Obs}}=I_{\text{Obs}}\omega$.  Using the rotation estimate of $\omega\sim 10^{-13}~\text{rad}/\text{yr}$ we obtain $J_{\text{Obs}} \sim 10^{87} \text{J}\cdot\text{s}$.  This is within a single order of magnitude of the calculated value of $J_0$ in Eq.~ (\ref{e:J0Lambda}).
Using the smallest and largest values of $\omega$ above as $\omega \sim 10^{-21}~\text{rad}/\text{yr}-10^{-9}~\text{rad}/\text{yr}$ gives us a range of plausible upper limits to a $J_{\text{Obs}}$ of the universe as $J_{\text{Obs}}\sim 10^{79}\text{J}\cdot\text{s}-10^{91}\text{J}\cdot\text{s}$. 
The cosmic angular momentum constant $J_0$ in~Eq.~(\ref{e:Lambda4}) clearly fits within this range, matching within several orders of magnitude of the estimated values.

\section{Fermions and Dark Matter}\label{s:Fermions}
We now briefly present how projective geometry enters into a discussion of fermions as a potential source for dark matter. Fields on the four-manifold are introduced into  the Lagrangian as scalars under projective transformations.  This follows since the Lie derivative of any $\l$ independent scalar has vanishing Lie derivative with respect to $\Upsilon$.  Also, the equi-projective extended vector fields that we use are of the form $A^\mu = \{A^0, \cdots A^3,0 \}$  and have a Lie derivative with respect to $\Upsilon $ that vanishes, i.e. 
\be \mathfrak{L}_\Upsilon A^\mu = \Upsilon^\a \partial_\a A^\mu -  A^\a \partial_\a \Upsilon^\mu =0.
 \ee  For projectively invariant fermions we need to compute their Lie derivative with respect to \(\Upsilon \). For the fermions  we will use the Kosmann derivative \cite{Kosmann:1966,Kosmann:1966B,Kosmann:1967,Kosmann:1971,Leao:2015} to determine the conditions on \( f(\ell)\)  so that  fermions transform trivially from the  Lie derivative with respect to \( \Upsilon \).      
The projective connection  acts on the gamma matrices via 
\be
\tilde{\nabla}_\mu \g^\nu = \partial_\mu  \g^\nu + [\tilde \Omega_\mu, \g^\nu] + \tilde{\G}^\nu_{\mu \sigma } \g^\sigma, 
\ee
and the spin connection on fermions is given by  
\be 
\tilde \Omega_\mu = \frac{1}{8} \tilde \o_{A B \mu} \g^A \g^B.
\ee
 In four dimensions, the fermion representation does not change when adding the $\g^4$.  Therefore, the projective connection on chiral fermions will introduce a natural axial coupling to projective gravity, as we will see shortly.

To continue with the Lie derivative, we have that for  a connection $\tilde{\G}^\b_{\,\,\m \a }$ and spinor connection $\tilde \O_\mu$, the Lie derivative of a spin $\frac{1}{2}$ field, $\psi$ with respect to a vector field $\beta^\a$ is given by
\be 
\mathfrak{L}_\b \psi = \b^\a \left(\partial_\a + \tilde\o_{A B \a} \g^A \g^B \right) \psi - \frac{1}{8}   \left(\tilde \nabla_\m \b_\n -\tilde \nabla_\n \beta_\m \right)\g^\m \g^\n  \psi.
\ee
Requiring that \(\mathfrak{L}_\Upsilon \psi=0,\) yields the condition 
\be 1- \frac{\l}{4} \frac{d}{d\l} \log(f(\ell))=0,
\ee
which implies that \( f(\ell) = (\frac{\l}{\l_0})^4 \).   With this the ``volume component'' of the spinor connection, \( \O_4 =0.\) Here we consider four component  Dirac fermions, $\psi^I$. Then the interaction  action for  the $I$th spinor \( \psi^I \) with mass \( M^I \) is given by (here there is no sum over $I$),
\be
S^I_{Dirac}= \int  \sqrt{|g|} f(\ell) d\ell\, d^4x\,\mathcal{L}^I_{Dirac},
\ee where 
\be\label{e:Dirac} 
 \mathcal{L}^I_{Dirac}=i \hbar\, c \,\bar\psi^I \g^\mu \tilde \nabla_\mu \psi^I  - c^2 M^I \bar\psi^I\psi^I = \bar\psi^I\left( i\hbar\, c\, \g^a \nabla_a -c^2 M^I-\hbar\, c\,\Phi \g^5\right)\psi^I.  
\ee
We observe there is an axial scalar coupling through,
\be \Phi =  \frac{1}{\l f(\ell)}  + \frac{\l}{4} \mathcal{D}=  \frac{1}{\l f(\ell)}  +\frac{\l}{4}(  \frac{mc}{J_0 } \phi  + \frac{4}{ 3} \Lambda) 
\ee
which due to the $\gamma^5$ in Eq.~(\ref{e:Dirac}) is CP violating.
The projective geometry has induced an axial scalar coupling to every fermion through \(\mathcal{D } \) and has generated a chiral asymmetric mass term, 
\be M^I_5= M^I +  m_{\ell \Lambda} \g^5=  
 \left(
\begin{array}{cccc}
 \text{M}^I & 0 & m_{\ell \Lambda} & 0 \\
 0 & \text{M}^I & 0 & m_{\ell \Lambda} \\
 m_{\ell \Lambda} & 0 & \text{M}^I & 0 \\
 0 & m_{\ell \Lambda}& 0 & \text{M}^I \\
\end{array}
\right),
\ee
where $ m_{\ell \Lambda}=  \frac{\hbar}{ c}\left( \frac{1}{\l f(\ell)}+ \frac{4}{ 3} \Lambda\right)$. After doing the $ \ell $ integral, with  $f(\ell) = \frac{\l^4}{\l_0^4}=\ell^4$ and where  $\ell_f$ is chosen so 
\[  \int_{\ell_i}^{\ell_f} d\ell f(\ell) = 1, \] the mass eigenvalues are 
\be m^I_\pm= M^I  \pm \frac{4 \hbar \Lambda}{3c} \pm  \frac{\hbar}{c \l_0}\log\left(\frac{(5+\ell_i^5)^{\frac{1}{5}}}{\ell_i}\right). \ee 
Thus the parameter $\ell_i$ tunes the axial contributions for the fermion masses.  The function, $  \frac{\hbar}{c \l_0}\log\left(\frac{(5+\ell_i^5)^{\frac{1}{5}}}{\ell_i}\right)$, is positive definite and cannot be used to set the $\l_0$ scale.  The parameters must be chosen so the total mass is non-negative.  The values of  $\l_0$ and $\ell_i$ will  be further constrained by phenomenology. This is presently being investigated.  It should be noted that  the axial scalar coupling also provides a portal for non-standard model fermions to interact with standard model fermions.  This will put further constraints on this axial scalar interaction.

\section{Conclusion}
We have further developed the Thomas-Whitehead gravitational theory and its phenomenological applications to dark energy and some issues related to dark matter.  
We have demonstrated that a $(\rd+1)$-dimensional action consisting of a pure projective Gauss-Bonnet term constructed out of projective curvature quantities naturally produces an Einstein-Hilbert term with cosmological constant and in $\rd=4$ introduces a new angular momentum constant, \( J_0 \),  of cosmological scale.  We gave a simple order of magnitude plausibility argument for what can be described as the angular momentum of the universe that is consistent with today's cosmological measurements and on the order of \( J_0 \). This arose from a natural decomposition of the diffeomorphism field in terms of non-dynamical degrees of freedom and dynamical degrees of freedom.  Furthermore, we were able to find the interaction of this field with fermions through the Dirac equation.  The theory predicts that fermion masses will receive an axial dependent contribution through the trace of the diffeomorphism field and the projective spin connection.   The interaction itself acts as a dark matter source as well as a portal for non-standard model fermions.   The phenomenological consequences of this are under investigation.  It should be noted that the origins of this theory are rooted in principles related to sprays\cite{Crampin} and projective Tractor calculus\cite{Eastwood} that are manifest in  Einstein geodesics and string theory.  The use of the projective Gauss-Bonnet action in four dimensions gives rise to dynamics for the diffeomorphism field without introducing higher derivative terms to the metric. We also give rationale for the absence of ghosts and tachyons in the scalar sector of the field theory.  This may be viewed as a covariant but non-linear strategy to include fluctuations to Einstein gravity.   
\section*{Acknowledgments}
The authors would like to thank Leo Rodriguez,  Shanshan Rodriguez, and members of the Nuclear and High Energy Theory group at the University of Iowa  for discussion.  We also thank Xiaole Jiang, Biruk Chafamo, Yehe Yan, Eric Peters, Alexis Leali, Patrick Vecera, Cole Dorman,  Salvatore Quaid, Taylor De Mello, and Indira Sheumaker from the summer research programs at Bates College,  the University of Iowa, and Grinnell College.      The research of K.\ S.\ is supported in part by the endowment of the Ford Foundation Professorship of Physics at Brown University.

\appendix
\section{Conventions and units}\label{a:Conventions}
The units of the various constants used throughout this paper for $\rd = 4$ are
\begin{align}
\begin{split}
        [J_0] =& \frac{M L^2}{T} ~~~,~~~[D_{ab}] = [\Lambda]= [R_{ab}] = L^{-2}~~~,~~~ [\alpha_0] = L^2   \cr
        [\lambda_0]  =& [a] = L ~~~,~~~[\ell] =[r] = [k]  =  \text{dimensionless}~~~,~~~[t]=T \cr
        \left[\kappa \right] =& \frac{T^2}{ML} ~~~,~~~[\rho] = \frac{M}{L^3}~~~,~~~[p]=\frac{M}{L T^2}~~~,~~~[H]=T^{-1}~~~,~~~[d^{\rd}x] = T L^{\rd -1}
        \end{split}
\end{align}
We may at times set $c=1$ but expose factors of $c$ when calculating numerical values.
Latin indices take values $a,b,\dots = 0,1,2,\dots, \rd-1$ and Greek indices take values \\ $\mu,\nu,\dots = 0,1,2,\dots, \rd$, with the exception of the Greek letter $\lambda$, which refers to the projective coordinate $x^{\rd} = \lambda = \lambda_0 \ell$.
Our conventions for the Riemann curvature tensor $R^{a}{}_{bcd}$ are the same as for the projective curvature $K^\mu{}_{\nu\alpha\b}$. The Riemann curvature tensor is written in terms of $\G^{m}{}_{ab}$ where as the projective curvature is written in terms of $\tilde{\G}^{\mu}{}_{\a\b}$:
\begin{align}\label{e:PCurvApp}
        K^\mu{}_{\nu\alpha\b} \equiv \tilde{\G}^\mu{}_{\nu[\b,\a]} + \tilde{\G}^\rho{}_{\n[\b}\tilde{\G}^\mu{}_{\a]\rho}~~~.
\end{align}
Here and throughout, brackets mean anti-symmetrization and parenthesis symmetrization. 
\begin{align}
K^{\alpha}{}_{\beta[\mu\nu]} =& K^{\alpha}{}_{\beta\mu\nu} - K^{\alpha}{}_{\beta\nu\mu} ~~~,~~~ K_{(\mu\nu)} = K_{\mu\nu} + K_{\nu\mu}
\end{align}
Eq.~(\ref{e:PCurvApp}) means the following must be true
\be \label{e:Kaction}
[{\tilde \nabla}_\a,{\tilde \nabla}_\b] V^\g = \,{K}^{\g}_{\,\,\,\rho\a \b  } V^\rho~~~,~~~[{\tilde \nabla}_\a,{\tilde \nabla}_\b] V_\g =- \,{K}^{\rho}_{\,\,\,\g\a \b  } V_\rho.
\ee
We define the $\rd$-dimensional Christoffel symbol $\G^{m}{}_{ab}$ in the usual way
\begin{align}
        \G^{m}{}_{ab} = \frac{1}{2} g^{mn}(g_{n(a,b)} - g_{ab,n})~~~,
\end{align}
but as $G_{\m\n}$ is not compatible with $\tilde{\G}^{\a}{}_{\m\n}$, the analogous definition for $\tilde{\G}^{\a}{}_{\m\n}$ is not correct. Instead, $\tilde{\G}^a{}_{mn}$ is defined in Eq.~(\ref{e:Gammatilde}). We define the projective curvature 3-tensor as
\begin{align}
        K_{\beta\mu\nu} \equiv K^{\alpha}{}_{\beta\mu\nu} \omega_\alpha = K^{\lambda}{}_{\beta\mu\nu}\lambda^{-1}~~~.
\end{align}
We contract over the first and fourth indices of the curvature tensor to form the Ricci tensor
\begin{align*}
        K_{\mu\nu} = K^{\alpha}{}_{\mu\nu\a}~~~.
\end{align*}
The $\rd$-dimensional metric $g_{ab}$ is embedded in the $(\rd+1)$-dimensional metric $G_{\alpha\beta}$, Eq.~(\ref{e:Gmetric}, \ref{e:GmetricInverse}),
\begin{align}
        G_{\a \b} =& \left( 
                        \begin{array}{cc}
                                g_{ab} & 0 \\
                                0 & -f(\ell)^2
                        \end{array}
        \right)~~~, \\
        G^{\a \b} =& \left( 
                        \begin{array}{cc}
                                g^{ab} & 0 \\
                                0 & -f(\ell)^{-2}
                        \end{array}
        \right)~~~.
\end{align}
where the $\rd$-dimensional metric $g_{ab}$ has signature $(+,-,-,-,\cdots,-)$ and the dimensionless parameter $\ell = \lambda/\lambda_0$. The $\rd$-dimensional Riemann Curvature tensor $?R^a_bcd?$ satisfies the same relation as the \text{$(\rd+1)$}-dimensional tensor $?K^\alpha_\beta\mu\nu?$, Eq.~(\ref{e:Kaction}), but in terms of the $\rd$-dimensional covariant derivative $\nabla_a$. The commutator of covariant derivatives on an arbitrary rank $m$-covariant, rank $n$-contravariant tensor is equivalent to the following action of $?R^a_bcd?$
\begin{align}\label{e:RactionGeneral}
        [\nabla_a , \nabla_b ] ?T_{c_1\dots c_m}^{d_1\dots d_n}? =& -?R^e_{c_1 ab}? ?T_{ec_2\dots c_m}^{d_1 d_2 \dots d_n}? - \dots -?R^e_{c_m ab}? ?T_{c_1 c_2\dots e}^{d_1 d_2 \dots d_n}? \cr
        &+ ?R^{d_1}_{eab}? ?T_{c_1\dots c_m}^{e\dots d_n}? + \dots + ?R^{d_m}_{eab}? ?T_{c_1\dots c_m}^{d_1\dots e}? 
\end{align}
Finally, we list all non-vanishing connections and curvatures below
\begin{align}
        \label{e:Gammaa}
        \tilde{\G}^\l{}_{ab} = \l \cD_{ab}~~~,~~~\tilde{\G}^a{}_{\l b}=&\tilde{\G}^a{}_{b \l} = \lambda^{-1} \delta_{b}{}^a~~~,~~~\tilde{\G}^a{}_{bc} = \G^a{}_{bc}~~~, \\
        \label{e:KRiemanna}
        K^a{}_{bcd} = R^a{}_{bcd} + \delta_{[c}{}^a \cD_{d]b}~~~,&~~~
        K^\l{}_{cab} = \lambda \partial_{[a} \cD_{b]c} + \lambda \Gamma^d{}_{c[b}\cD_{a]d}~~~, \\
        \label{e:KRiccia}
        K_{ab} = K^{\mu}{}_{ab\mu} = & R_{ab} -(\rd - 1) \cD_{ab} \\
        \label{e:Ka}
        K\equiv & G^{\alpha\beta}K_{\alpha\beta} = R - (\rd - 1)\cD \\
        \label{e:RDa}
                R =& g^{ab}R_{ab}~~~,~~~ \cD =  g^{ab}\cD_{ab}~~~,
\end{align}
along with the tensor decomposition of $\mathcal{D}_{ab}$
\begin{align}\label{e:ShiftA}
        \cD_{ab} = &D_{ab} + \frac{4}{\rd (\rd - 1)} \Lambda g_{ab} \cr
               = &\left[W_{ab} + \frac{mc}{J_0 \rd}g_{ab} \phi \right] + \frac{4}{\rd (\rd - 1)} \Lambda g_{ab} ~~~
\end{align}
and the relation between $D_{ab}$ and curvature
\begin{align}
\label{e:tDstara}
        \tD_*^{ab} =& (\rd -1 ) g^{ab} \tD - 2 (2 \rd- 3)\tD^{ab}\\
        \label{e:tDaba}
        \tD_{ab} =& (\rd - 1)D_{ab} -2 R_{ab} = -K_{ab} - R_{ab}\ - \tfrac{4}{\rd}\Lambda g_{ab}\\
        \label{e:tDa}
        \tD = g^{ab}\tD_{ab} =& (\rd - 1)D - 2 R = -K - R - 4 \Lambda~~~.
\end{align}
We note that $W_{ab}$ is traceless $W_{ab}g^{ab} = 0$ and write the general star $(*)$ operator used throughout the paper
\begin{align}\label{e:Stara}
        T_*^{ab} =& M^{abmn} T_{mn} = (\rd - 1) g^{ab} T - 2 (2 \rd - 3) T^{ab} 
\end{align}
where $T = g^{mn}T_{mn}$ and the $M^{abmn}$ tensor is
\begin{align}\label{e:Mtensora}
        M^{abmn} =& (\rd -1) g^{ab}g^{mn} - 2 (2 \rd - 3) g^{am}g^{bn}~~~.
\end{align}

\section{General Relativity and Cosmology Review} \label{a:Cosmo}
Here we present a quick proof of Einstein's field equations from the Einstein-Hilbert action and a brief overview of standard cosmology in four space-time dimensions.  In the following the constants $s_i$ are convention dependent and are equal to plus or minus one.  The various conventions in the literature are given in table~\ref{tab:GRconventions}
\begin{table}[!h]
\centering
\begin{tabular}{|c|c|c|c|c|c|c|}
   \hline
       Reference & $s_1$ & $s_2$ & $s_3$ & $s_4$  \\
   \hline
       Kolb \& Turner~\cite{kolbt} & - & +  & + & -  \\ 
   \hline
      MTW~\cite{Misner:1974qy}, Liddle \& Lyth~\cite{Liddle:2000}  & + & +  & + & + \\
   \hline
      HEL~\cite{Hobson:2006se} & - & +  & - & ~ \\
\hline
      Weinberg~\cite{Weinberg:2008zzc} & + & ~  & - & - \\
\hline   
  Weinberg~\cite{Weinberg:1972}, RY~\cite{Rodgers:2006ep} & + & -  & - & - 
\\      
\hline
Dirac~\cite{Dirac:1975} & - & +   & - & -  \\
\hline
Ohanian \& Ruffini\cite{Ohanian:1995uu} & - & +  & - & +  \\
\hline
\end{tabular}
\caption{Sign conventions of different authors.}\label{tab:GRconventions}
\end{table}
In this paper we use the conventions of Ohanian \& Ruffini~\cite{Ohanian:1995uu}.  Also, in the ``mathtensor" package of Mathematica, the default setting are: 
$s_2 = \text{Rmsign} = +1,~s_3 = s_2 \text{Rcsign} = +1,~s_1 = \text{MetricgSign} = +1$. 

The cosmological principle demands the large scale structure of the universe to be spatially homogeneous and isotropic.  The metric encompassing these qualities is known as the Friedmann-Lemaitre-Robertson-Walker (FLRW) metric
\begin{align}\label{e:FRW}
   s_1 g_{mn} dx^m dx^n &= -dt^2 + a(t)^2\left(\frac{dr^2}{1 - k r^2} + r^2( d\theta^2 +  \sin^2\theta \, d\phi^2)\right)
\end{align}
The stress tensor and equations of motion for the diffeomorphism field derived in this paper assume a metric of constant volume $\partial_a \sqrt{g} = \sqrt{g} \Gamma^b{}_{ab} = 0$. Therefore, the vacuum solution presented in section~\ref{s:Solutions} is in terms of a constant volume FLRW metric that is a coordinate transformation of Eq.~(\ref{e:FRW}). An example of such a coordinate transformation for $k=0$ is $a^3(t) dt = d\tau$, $x = r \sin\theta\cos\phi$, $y = r \sin\theta\sin\phi$, and $z = r\cos\theta$. A constant volume FLRW metric with $k=\pm 1$ can be found as well.

The Riemann curvature tensor and Ricci tensors can be defined independent of convention as
\begin{align}
    s_2 ?R^a_mbn? &= -?\Gamma^a_mb,n? + ?\Gamma^a_mn,b? - ?\Gamma^c_mb??\Gamma^a_nc? + ?\Gamma^c_mn??\Gamma^a_bc?, \\
    s_3 R_{mn} &= s_2 ?R^a_man? = -?\Gamma^a_ma,n? + ?\Gamma^a_mn,a? - ?\Gamma^c_ma??\Gamma^a_nc? + ?\Gamma^c_mn??\Gamma^a_ac?.
\end{align}
so that $s_3$ is the sign of the curvature of a sphere.  The Christoffel symbol is given in terms of the metric by
\begin{align}
    ?\Gamma^a_mn? &= \frac{1}{2} g^{ab}(g_{bm,n} + g_{bn,m} - g_{mn,b})
\end{align}

Defining the Ricci scalar as $ R = R_{mn} g^{mn}$, the Einstein Equations are derived from the Einstein-Hilbert action plus source $S_{source}$
\begin{align}
    S &= \frac{s_4 }{2\kappa}\int d^4x \sqrt{|g|} (R - 2 s_1 s_3 \Lambda) + S_{source}
\end{align}
where $\kappa= 8 \pi G/c^4$  and $\Lambda$ is the cosmological constant.
Variation of the action yields
\begin{align}
     \delta_g S = 0 = \int d^4x \sqrt{|g|} &~\delta g^{mn} \left(R_{mn}- \frac{1}{2} g_{mn} R + s_1 s_3 \Lambda g_{mn} -  s_3 \kappa \Theta_{mn} \right) + \nonumber\\
&+ \int d^4x \sqrt{|g|} \delta_g R_{mn} g^{mn}
\end{align}
where we have defined the stress-energy tensor $\Theta_{mn}$ through
\begin{align}
     \delta_g S_{source} &= \frac{s_3 s_4}{2} \int d^4x \sqrt{|g|} \Theta^{mn}\delta g_{mn} = -\frac{s_3 s_4}{2} \int d^4x \sqrt{|g|} \Theta_{mn}\delta g^{mn}.
\end{align}

We can discard the last term in the action's variation as it yields the surface term
\begin{align}
    \int d^4x \sqrt{|g|} \delta_g R_{mn} g^{mn} &= \int d^4x ( \sqrt{|g|} g^{mn} \delta? \Gamma^a_m[n ?)_{;a]} = 0
\end{align}
where $;$ denotes a covariant derivative.
As promised, Einstein's Equations become
\begin{align}
    G_{mn} &\equiv R_{mn} - \frac{1}{2}g_{mn} R =  s_3\kappa \Theta_{mn} - s_1 s_3 \Lambda g_{mn}
\end{align}

 With the FLRW metric, Eq.~(\ref{e:FRW}), and a stress tensor of the form for a perfect fluid
\begin{align}
   \Theta_{mn} &= (\rho + p ) ?\delta_m^0? ?\delta_n^0? + s_1 g_{mn} p,
\end{align}
with $\rho$ the mass density and $p$ the pressure of the Universe, the Einstein equations become what are known as the Friedmann equations:
\begin{align}\label{e:FriedmannH}
    H(t)^2 &= \frac{\rho}{3 \kappa^{-1}}  + \frac{\Lambda}{3}  - \frac{k}{a^2},~~~(00~\mbox{equation of motion}), \\
\label{eq:EEqij}  
\frac{\ddot{a}}{a} &= - \frac{\kappa}{6}(\rho + 3 p) + \frac{\Lambda}{3} ,~~~(ij~\mbox{eqm. with 00 eqm.)}
\end{align}
where the Hubble parameter is
\begin{align}
   H(t) \equiv \frac{\dot{a}(t)}{a(t)}.
\end{align}

Notice a positive cosmological constant will accelerate the scale factor, $a(t)$, as evidenced in Eq.~(\ref{eq:EEqij}) that it has the opposite sign as pressure. In this way, the cosmological constant, or presumably dark energy which is its cause, acts like a \emph{negative} pressure tending to pull the universe apart rather than squeeze it together as one would expect from a regular, positive pressure.  These Friedmann equations are redundant with the continuity equation
\begin{align}
     \nabla_{n} \Theta^{mn} &= 0 \to \left\{ \begin{array}{ll}
         \dot{\rho} + 3 H (\rho + p) = 0\\
         p = p(t)
\end{array}
\right.
\end{align}
which is actually sign convention independent with the form of the perfect fluid given above.
The system can then be succinctly described by the either the $ij$ equation or $00$ equation of motion and the continuity equation. We define the mass density and pressure of the vacuum ($\rho_\Lambda$,$p_\Lambda$)  and curvature ($\rho_k$,$p_k$)  as 
\begin{align}\label{e:rhopLambda}
        \rho_\Lambda =& - p_\Lambda = \kappa^{-1} \Lambda~~~\\
        \label{e:rhopk}
        \rho_k =& -3 p_k = -\frac{ 3 }{\kappa a^2} k~~~.
\end{align}
and combine them with $\rho$ and $p$ to form $\rho_c$ and $p_c$, respectively
\begin{align}
        \rho_c =& \rho + \rho_\Lambda + \rho_k \\
        p_c =& p + p_\Lambda + p_k~~~.
\end{align}
The quantity $\rho_c$ is known as the critical density as it is the critical value $\rho$ takes in a flat Universe ($k=0$) with no cosmological constant. The Friedmann equations can be succinctly written in terms of $\rho_c$ and $p_c$:
\begin{align}\label{e:FriedmannHsuccinct}
        \rho_c =& \frac{3}{\kappa} H^2 \\
        \label{e:Friedmannasuccinct}
        \frac{\ddot{a}}{a} =& - \frac{\kappa}{6} (\rho_c + 3 p_c)~~~.
\end{align} The pressure $p$ and mass density $\rho$ are a combination of contributions from matter ($p_m$, $\rho_m$), radiation ($p_r$, $\rho_r$), and any other source ($p_\text{other}$, $\rho_\text{other}$) such as the diffeomorphism field presented in this paper so we write
\begin{align}
                \label{e:p}
                p_c =& p_m + p_r  + p_\text{other} + p_\Lambda + p_k\\
                \label{e:rho}
                \rho_c =& \rho_m + \rho_r  + \rho_\text{other} + \rho_\Lambda + \rho_k~~~.    
\end{align} 
Note that for radiation or other massless fields, the mass density is defined as the energy-density per unit $c^2$:
\begin{align}
        \rho_r \equiv u_r/c^2~~~,~~~\text{and similar for other massless fields}.
\end{align}

There will in general be field equations to satisfy for the cosmological sources of $\rho$ and $p$ as well, such as the field equations for the diffeomorphism field in this paper. The diffeomorphism field equations and stress tensor derived in this paper are  in a gauge where the metric has constant volume.  Thus using these equations and stress tensor for the diffeomorphism field requires the Friedmann equations to be expressed in terms of a constant volume metric as well as described in the text after Eq.~(\ref{e:FRW}).

More generally, cosmological measurements of each species (matter, vacuum, etc.) are typically quoted in terms of a density parameter $\Omega_i = \rho_i/\rho_c$ for each species $i$: $i=m$ for matter, $i=\Lambda$ for vacuum (cosmological constant), etc. For instance, the density parameter for the vacuum  is defined as
\begin{align}\label{e:OmegaLambda}
        \Omega_\Lambda = \frac{\rho_\Lambda}{\rho_c}~~~.
\end{align}  
Eq.~(\ref{e:p}) is often written in terms of  $\Omega$:
\begin{align}
        1 = \Omega_m + \Omega_r + \Omega_{\text{other}} + \Omega_\Lambda + \Omega_k~~~.
\end{align}
Solving Eqs.~(\ref{e:OmegaLambda}), (\ref{e:FriedmannHsuccinct}), and~(\ref{e:rhopLambda}) for $\Lambda$ and putting in appropriate factors of the speed of light $c$ yields
\begin{align}\label{e:Lambda}
        \Lambda = 3 H^2 \Omega_\Lambda/c^2~~~.
\end{align}
The value of the Hubble parameter at $t = \text{today}$ is denoted as $H_0$. Recent measurements of $\Omega_\Lambda$ and $H_0$ by \emph{Planck} are~\cite{Akrami:2018vks} 
\begin{align}
        \Omega_\Lambda \approx 0.68~~~,~~~H_{0,\text{Planck}}\approx 67~\text{km}/\text{s}/\text{Mpc}~~~.
\end{align}
On the other hand, the Reiss collaboration of cosmic distance ladder redshift measurements finds the following measurement of the Hubble Parameter~\cite{Riess:2019cxk}:
\begin{align}
H_{0, \text{Riess}} \approx 74~\text{km}/\text{s}/\text{Mpc}~~~.        
\end{align}
These two measurements are both too precise to be in agreement with each other, a problem known as the Hubble tension. For all calculations in this paper, we thus take the Hubble parameter to be an average of the two measurements and use Planck's measurement of $\Omega_\Lambda$
\begin{align}\label{e:OmegaH}
        \Omega_\Lambda \approx 0.68~~~,~~~H_{0}\approx 71~\text{km}/\text{s}/\text{Mpc}~~~.
\end{align}
Plugging these values into Eq.~(\ref{e:Lambda}) yields the measured value of the Cosmological constant we will use throughout the paper.
\begin{align}
        \Lambda \approx 1.2 \times 10^{-52} \text{m}^{-2}~~~.
\end{align}
Often, the Cosmological constant is given in terms of its associated mass density, Eq.~(\ref{e:rhopLambda}). For the above value of the cosmological constant, this density is
\begin{align}
        \rho_\Lambda =&\frac{c^2}{8 \pi G} \Lambda = 5.9 \times 10^{-27} \text{kg}/\text{m}^3
\end{align}
where we have used Eq.~(\ref{e:kappasquared}) to write out $\kappa^{-1}$ in terms of $G$ and put back factors of $c$.
Written in natural units, this is
\begin{align}
        \rho_\Lambda =& \frac{\hbar^3 c^7}{8 \pi G}  \Lambda = 2.5 \times 10^{-47} \text{GeV}^4 
\end{align}


\section{Expansion of the Projective Gauss-Bonnet Lagrangian}\label{a:Ldetails}
We expand the projective Gauss-Bonnet Lagrangian $\mathcal{L}_{TW}$ in terms of the diffeomorphism field and $\rd$-dimensional curvature tensors as follows. Using Eq.~(\ref{e:K}), the projective curvature scalar squared $K^2$ is
\begin{align}\label{e:KExplicit}
        K^2     =& (R - (\rd - 1)\cD)^2 \cr
                =& R^2 + (\rd - 1)^2\cD^2 - 2(\rd - 1)R\cD \cr
                =& R^2 + (\rd - 1)\cD\tilde{\cD}~~~,
\end{align}
where we are introducing $\tilde{\cD}_{ab}$, defined as
\begin{align}
\widetilde{\cD} \equiv g^{ab}\widetilde{\cD}_{ab} ~~~,~~~\widetilde{\cD}_{ab} =& (\rd - 1)\cD_{ab} - 2 R_{ab}~~~.
\end{align}
Next, we calculate the projective Ricci squared $K_{\alpha\beta}K^{\alpha\beta}$:
\begin{align}\label{e:KRicciExplicit}
        K_{\alpha\beta}K^{\alpha\beta} =& K_{ab}K^{ab} + K_{\lambda\lambda}K^{\lambda\lambda} + 2 K_{a\lambda}K^{a\lambda} \cr
                =& (R_{ab} - (\rd - 1)\cD_{ab})(R^{ab} - (\rd - 1)\cD^{ab}) \cr
                =& R_{ab}R^{ab} + (\rd - 1)^2 \cD_{ab}\cD^{ab} -2 (\rd - 1) \cD_{ab}R^{ab} \cr
                =& R_{ab}R^{ab} + (\rd - 1)\cD_{ab}\widetilde{\cD}^{ab}~~~.
\end{align}
To calculate the projective Riemann curvature squared $K_{\alpha\beta\mu\nu}K^{\alpha\beta\mu\nu}$, we utilize our knowledge of the non-vanishing terms in Eq.~(\ref{e:KRiemann}) to first write
\begin{align}
        K_{\alpha\beta\mu\nu} = \delta_{\alpha}{}^\lambda G_{\lambda\lambda}K^\lambda{}_{\beta\mu\nu} + \delta_\alpha{}^a K_{a\beta\mu\nu}~~~,~~~K^{\alpha\beta\mu\nu} = \delta_b{}^\beta \delta_m{}^\mu \delta_n{}^\nu K^{\alpha bmn}~~~.
\end{align}
With this, we calculate $K_{\alpha\beta\mu\nu}K^{\alpha\beta\mu\nu}$ as
\begin{align}\label{e:KRiemannExplicit}
        K_{\alpha\beta\mu\nu}K^{\alpha\beta\mu\nu} =&  (
        \delta_{\alpha}{}^\lambda G_{\lambda\lambda}K^\lambda{}_{\beta\mu\nu} + \delta_\alpha{}^a K_{a\beta\mu\nu})\delta_b{}^\beta \delta_m{}^\mu \delta_n{}^\nu K^{\alpha bmn} \cr
        =& K^\lambda{}_{bmn}K^{\lambda bmn}G_{\lambda\lambda} + K_{abmn}K^{abmn} \cr
        =& \lambda^2 K_{bmn}K^{bmn} G_{\lambda\lambda} + (R_{abmn} + g_{a[m}\cD_{n]b})(R^{abmn} + g^{a[m}\cD^{n]b}) \cr
        =& -\lambda^2 f^2 K_{bmn}K^{bmn} + R_{abmn}R^{abmn} + 4 R^m{}_{bmn}\cD^{nb} \cr
        &+ 2 g_{am}\cD_{nb}g^{am}\cD^{nb} - 2 g_{an}\cD_{mb}g^{am}\cD^{nb} \cr
        =& -\lambda^2 f^2 K_{bmn}K^{bmn} + R_{abmn}R^{abmn}  + 2 \rd \cD_{ab}\cD^{ab} - 2 \cD_{ab}\cD^{ab} - 4 R_{ab}\cD^{ab} \cr
        =& -\lambda^2 f^2 K_{bmn}K^{bmn} + R_{abmn}R^{abmn}   + 2 (\rd - 1)\cD_{ab}\cD^{ab}  - 4 \cD_{ab}R^{ab}\cr
        =& -\lambda^2 f^2 K_{bmn}K^{bmn} + R_{abmn}R^{abmn} + 2 \cD_{ab}\tcD^{ab}
\end{align}
With the results of Eqs.~(\ref{e:KExplicit}), (\ref{e:KRicciExplicit}), and (\ref{e:KRiemannExplicit}), the projective Gauss-Bonnet Lagrangian becomes
\begin{align}
        \mathcal{L}_{TW} =& K^2 - 4K_{\alpha\beta}K^{\alpha\beta} + K_{\alpha\beta\mu\nu}K^{\alpha\beta\mu\nu} \cr
        =& R^2  + (\rd - 1) \cD\tcD - 4 (R_{ab}R^{ab} + (\rd - 1) \cD_{ab}\tcD^{ab}) \cr
        &- \lambda^2 f^2 K_{bmn}K^{bmn} + R_{abmn}R^{abmn} + 2 \cD_{ab}\tilde{\cD}^{ab}\cr
        =& \mathcal{L}_{GB} -\lambda^2 f^2 K_{bmn}K^{bmn} + (\rd - 1)\cD\tcD - 2 (2\rd-3) \cD_{ab}\tcD^{ab} 
\end{align}
where the d-dimensional Gauss-Bonnet Lagrangian is
\begin{align}
        \mathcal{L}_{GB} =& R^2 - 4 R_{ab}R^{ab} + R_{abmn}R^{abmn} ~~~.
\end{align}
We collect the terms quadratic in $\cD_{ab}$ and $\tcD_{ab}$ and use Eq.~(\ref{e:Star}) to define
\begin{align}
        \tcD_*^{ab} =& M^{abmn}\tcD_{mn} = (\rd -1)g^{ab} \tcD - 2 (2\rd-3)\tcD^{ab}
\end{align}
This  allows us to simplify the Lagrangian to
\begin{align}
        \mathcal{L}_{TW} =& \mathcal{L}_{GB} -\lambda_0^2 \ell^2 f^2 K_{bmn}K^{bmn}  + \cD_{ab}\tcD_*^{ab}~~~.
\end{align}
Introducing the coupling constant $J_0$ and a factor of $c$ for proper units, we construct the full action
\begin{align}
        S_{TW} =&- \tfrac{J_0 c}{2} \int d^\rd x d\ell \sqrt{|G|} \mathcal{L}_{TW} \cr
        =&  - \tfrac{J_0 c}{2} \int d^\rd x \sqrt{|g|} \int d\ell f(\ell) \mathcal{L}_{TW} \cr
        =& - \tfrac{J_0 c}{2}\int d^\rd x \sqrt{|g|} \int d\ell f(\ell) \Big(\mathcal{L}_{GB} - \lambda^2 f(\ell)^2 K_{mab}K^{mab}
         + \cD_{ab}\tcD_*^{ab} \Big) \cr
        =& - \tfrac{J_0 c }{2}\int d^\rd x \sqrt{|g|} \Big( \mathcal{L}_{GB} 
         + \cD_{ab}\tcD_*^{ab} \Big) \int_{\ell_i}^{\ell_f} d\ell  f(\ell)  \cr
        &+ \tfrac{J_0 c}{2}\int d^\rd x \sqrt{|g|}  K_{mab}K^{mab} \lambda_0^2\int_{\ell_i}^{\ell_f} d\ell \ell^2 f(\ell)^3   
\end{align}
where we have substituted $\lambda = \lambda_0 \ell$ and factored terms involving $\cD_{ab}$, $\tcD_*^{ab}$, and $K_{mab}$ out of the $\ell$ integral as these terms are $\ell$-independent. We have also introduced cutoff's $\ell_i$ and $\ell_f$. We can define one of these integrals to be whatever number we wish by ensuring $f(\ell)$ is properly normalized. The other integral, we will define as a new constant to be determined once $f(\ell)$ is chosen and appropriately normalized. With this in mind, we define
\begin{align}
        \int_{\ell_i}^{\ell_f} d\ell  f(\ell) = 1~~~,~~~\alpha_0 =& \lambda_0^2 \int_{\ell_i}^{\ell_f} d\ell \ell^2 f(\ell)^3   ~~~.
\end{align}
With these definitions, the $TW$ action becomes
\begin{align}
        S_{TW}=& -\tfrac{J_0 c}{2}\int d^\rd x \sqrt{|g|} \Big(  \mathcal{L}_{GB} + \cD_{ab}\tcD_*^{ab} \Big) + \tfrac{J_0 c}{2} \int d^\rd x \sqrt{|g|}  ~\alpha_0 K_{mab}K^{mab}  
\end{align}
Simplifying we have
\begin{align}\label{e:STWapp}
    S_{TW} =& \int d^\rd x\sqrt{|g|} \mathcal{L}_\cD + S_{GB}~~~, \\
    \mathcal{L}_\cD = & \tfrac{J_0 c}{2}\left[\alpha_0  K_{bmn}K^{bmn} -  \cD_{ab}\tcD_*^{ab} \right] ~~~,  \\
    S_{GB} = & -\tfrac{J_0c}{2} \int d^\rd x \sqrt{|g|} \left( R^2 - 4 R_{ab}R^{ab} + R_{abmn}R^{abmn} \right)~~~
\end{align}
Next, we decompose as in Eq.~(\ref{e:Shift})
\begin{align}
        \cD_{ab} = &D_{ab} + \frac{4}{\rd (\rd - 1)} \Lambda g_{ab}
\end{align}
This results in the following decomposition for $\tcD_*^{ab}$
\begin{align}
        \tcD_*^{ab} = \tD_*^{ab} + \tfrac{1}{2}y(\rd) \Lambda g^{ab}
\end{align}
with $y(\rd)$ as in Eq.~(\ref{e:yd}) and
$K_{abc} = \nabla_b \cD_{ca} - \nabla_{c} \cD_{ba}$ unchanged because of the covariance of the metric. The Lagrangian $\mathcal{L}_\cD$ becomes
\begin{align}\label{e:LDplusDeltaLD}
        \mathcal{L}_\cD=& \tfrac{J_0 c}{2}\left[\alpha_0  K_{bmn}K^{bmn} -  D_{ab}\tD_*^{ab} \right] + \Delta\mathcal{L}_D 
\end{align}
where
\begin{align}
        \Delta \mathcal{L}_D =& -\tfrac{J_0 c}{2}\left[  \frac{4}{\rd (\rd -1)} \Lambda \tD_* + \tfrac{1}{2} y(\rd) \Lambda D + \frac{2y(\rd)}{\rd -1} \Lambda^2\right]
\end{align}
Using Eq.~(\ref{e:Star}) and~(\ref{e:tD}), we rewrite $\tD_*$ as
\begin{align}
        \tD_* =& \frac{\rd (\rd - 1)}{8}y(\rd) D - \frac{\rd \,y(\rd)}{4} R
\end{align}
resulting in the following for $\Delta\mathcal{L}_D$
\begin{align}
        \Delta \mathcal{L}_D =& \frac{1}{2} \frac{y(\rd)}{\rd-1} \Lambda J_0 c  \left[ R  - 2\Lambda -(\rd - 1) D )\right]
\end{align}
Thus, producing the Einstein-Hilbert term with the correct coefficient demands
\begin{align}\label{e:kappabeta}
        \kappa =& \frac{\rd-1}{y(\rd)\Lambda J_0 c } 
\end{align}
so that
\begin{align}\label{e:DeltaLD}
        \Delta \mathcal{L}_D =& \frac{1}{2\kappa}  \left[ R  - 2\Lambda -(\rd - 1) D )\right]~~~.
\end{align}
Substituting this into Eq.~(\ref{e:LDplusDeltaLD}) and integrating results in the first two terms of the action 
\begin{align}\label{e:STWapp1}
    S_{TW} =& \frac{1}{2\kappa} \int d^\rd x\sqrt{|g|} \left( R - 2 \Lambda\right)+ \int d^\rd x\sqrt{|g|} \mathcal{L}_D + S_{GB}~~~, \\
    \label{e:LDapp1}
    \mathcal{L}_{D} 
    = & \tfrac{J_0 c}{2}\left[\alpha_0 K_{bmn}K^{bmn} - D_{ab}\tD_*^{ab} - y(\rd) \Lambda D \right]~~~
\end{align}
where $S_{GB}$, Eq.~(\ref{e:SGBa}), has come along for the ride.

\section{Equations of Motion Derivation}\label{a:EQMDerivation}
In $S_{TW}$, only $\mathcal{L}_D$ contains the diffeomorphism field $D_{ab}$. Therefore, we restrict our derivation of the diffeomorphism field equations of motion to the variation of $\mathcal{L}_D$:
\begin{align}
        -\frac{2}{J_0 c}\delta \mathcal{L}_D = & -2\alpha_0 K^{mab} \delta K_{mab} + ( \delta D_{ab}\tD_*^{ab} + D_*^{ab} \delta\tD_{ab}) + y(\rd) \Lambda g^{ab}\delta D_{ab}
        \cr
        =& -2\alpha_0K^{mab} \nabla_{[a} \delta D_{b]m}   +  \left[\tD_*^{ab} + (\rd -1)D_*^{ab} + y(\rd) \Lambda g^{ab}\right]\delta D_{ab} \cr
        =& -4\alpha_0 K^{mab} \nabla_{a} \delta D_{bm}   -2 K_*^{ab}\delta D_{ab} 
\end{align}
In the last line we have used the fact that $K^{mab}$ is antisymmetric in its last two indices as well as Eqs. (\ref{e:tDab}), (\ref{e:Star}), and (\ref{e:yd}) to make the substitution
\begin{align}
 \tD_*^{ab} + (\rd - 1) D_*^{ab} + y(\rd)\Lambda g^{ab} = - 2 K_*^{ab} ~~~.
\end{align} 
 After rewriting the derivative term up to a total derivative, renaming indices, and simplifying, the variation $\delta \mathcal{L}_D$ becomes
\begin{align}
        -\frac{2}{J_0 c}\delta \mathcal{L}_D =&\left[ -4 \alpha_0 \nabla_n K^{abn} - 2K_*^{ab}  \right]  \delta D_{ab}~~~.
\end{align}
Symmetrizing over $ab$ and setting the factor in parenthesis to zero yields the diffeomorphism equation of motion
\begin{align}
        \alpha_0 \nabla_n K^{(ab)n}  =& -K_*^{ab} ~~~.
\end{align}

\section{Stress-Energy Tensor Derivation}\label{a:StressEnergyDerivation}
The stress-energy tensor for a source action $S_{source}$ is defined as
\begin{align}\label{e:ThetaDef}
     \delta S_{source} &= -\tfrac{1}{2} \int d^\rd x \sqrt{|g|} \Theta^{ab}\delta g_{ab} = \tfrac{1}{2} \int d^\rd x \sqrt{|g|} \Theta_{ab}\delta g^{ab}.
\end{align}
For action variations involving terms such as the following containing $\delta \Gamma^c{}_{ab}$
\begin{align}
        \tV_c{}^{ab}\delta \Gamma^c{}_{ab}
\end{align}
with $\tV_c{}^{ab}$ symmetrized in its last two indices
\begin{align}
        \tV_c{}^{ab} = \tV_c{}^{ba}
\end{align}
we shall find the following useful in deriving the associated stress-energy tensor:
\begin{align}\label{e:varSGamma}
        \delta S_{source} =& -\tfrac{1}{2} \int d^\rd x \sqrt{|g|} \Theta^{ab}\delta g_{ab}  \cr
                                                =& -\tfrac{1}{2} \int d^\rd x \sqrt{|g|} \Theta^{ab}_{(g)}\delta g_{ab}   + \int d^\rd x \sqrt{|g|} \tV_c{}^{ab} \delta\Gamma^{c}{}_{ab}  \cr
                                                =&  -\tfrac{1}{2} \int d^\rd x \sqrt{|g|} \left[ \Theta^{ab}_{(g)} + \nabla_c \tV^{(ab)c} - \nabla_c \tV^{cab} \right]\delta g_{ab}   ~~~.
\end{align}
Comparing the first and last lines of Eq.~(\ref{e:varSGamma}) we conclude that the full stress tensor for $S_{source}$ is
\begin{align}
        \Theta^{ab} =& \Theta^{ab}_{(g)} + \nabla_c \tV^{(ab)c} - \nabla_c \tV^{cab} 
\end{align}
where $\Theta^{ab}_{(g)}$ is the term associated with $\delta g_{ab}$, and not $\tV_c{}^{ab}$, as in the second and third lines of Eq.~(\ref{e:varSGamma}). In the next section we will prove Eq.~(\ref{e:varSGamma}).  For now we will simply use it to derive the stress-energy tensor associated with $\mathcal{L}_D$, Eq.~(\ref{e:LD}). In doing so, we will find the following property useful
\begin{align}\label{e:useful1}
        \delta g^{ab} Y_{ab} = - \delta g_{ab} Y^{ab} 
\end{align}
for some arbitrary tensor $Y_{ab}$. Also, the mixed rank tensor $?M_c^b_d^n?$ is
\begin{align}\label{e:useful2}
        ?M_c^b_d^n? =& g_{cp}g_{dm} M^{pbmn}= (\rd - 1)\delta_c{}^b\delta_d{}^n - 2 (2m-3) \delta_c{}^n\delta_d{}^b ~~~.
\end{align}
We use this last definition to write the source Lagrangian in the most useful form for our present purposes as below. Neglecting the Gauss-Bonnet portion, as we will ultimately focus on $\rd =4$, the source action is
\begin{align}
    S_D =& \int d^\rd x\sqrt{|g|} \mathcal{L}_D ~~~, \\
    \label{e:LDapp}
    \mathcal{L}_{D} 
    = & \tfrac{J_0 c}{2}\left[\alpha_0 K_{bmn}K_{apq}g^{ab}g^{mp}g^{nq} - D_{ab}\tD_{mn}g^{ac}g^{md} ?M_c^b_d^n? - y(\rd) \Lambda D_{ab} g^{ab} \right]~~~.
\end{align}
To derive the associated stress-energy tensor, we vary this with respect to the metric
\begin{align}
   \delta S_D = &       \int d^\rd x \left[\sqrt{|g|} \delta \mathcal{L}_D + \tfrac{1}{2}\sqrt{|g|}g^{ab} \delta g_{ab} \mathcal{L}_D  \right] \cr
   =& - \tfrac{1}{2}\int d^\rd x \sqrt{|g|}\left[ -2 \delta \mathcal{L}_D- g^{ab} \mathcal{L}_D  \delta g_{ab}  \right] 
\end{align}
We now focus on the $\delta \mathcal{L}_D$ term. After a little simplification, this results in
\begin{align}
\begin{split}
        \tfrac{2}{J_0 c}\delta \mathcal{L}_D =& 2 \alpha_0 K^{bmn} \delta K_{bmn} + \alpha_0 K_{bmn} K_a{}^{mn} \delta g^{ab} + 2 \alpha_0 K_{bmn} K^{bm}{}_q \delta g^{nq}\cr
        &  -D_*^{mn} \delta \tD_{mn}- D_{ab} \tD_{mn} M_c{}^{bmn} \delta g^{ac} - D_{ab}\tD_{mn} M^{ab}{}_d{}^n \delta g^{md} \cr
        &- y(\rd) \Lambda D_{ab} \delta g^{ab}~~~.
        \end{split}
\end{align}
Expanding out the definition of $K_{bmn}$ and $\tD_{mn}$ and relabeling indices and using Eq.~(\ref{e:useful1}) to rewrite the variation in terms of the covariant $\delta g_{ab}$ we have
\begin{align}
\begin{split}
        -\tfrac{2}{J_0 c}\delta \mathcal{L}_D =& -4\alpha_0 K^{bmn} \delta (\nabla_m D_{nb}) + \alpha_0 K^{a}{}_{mn} K^{bmn} \delta g_{ab} + 2 \alpha_0 K^{mna} K_{mn}{}^b \delta g_{ab}\cr
        &  +D_*^{mn} \delta (-2 R_{mn})- ?D^a_c? \tD_*^{bc} \delta g_{ab} - D_*^{bc}?\tD^a_c? \delta g_{ab}- y(\rd) \Lambda D^{ab} \delta g_{ab}~~~.
        \end{split}
\end{align}
Next, we expand out the covariant derivative on $D_{nb}$ and use the fact that
\begin{align}
        \delta R_{mn} = \nabla_{[n} \delta ?\Gamma^c_c]m?
\end{align}
to simplify the variation to
\begin{align}
\begin{split}
        -\tfrac{2}{J_0 c}\delta \mathcal{L}_D =&  \left[\alpha_0 K^{a}{}_{mn} K^{bmn}  + 2 \alpha_0 K^{mna} K_{mn}{}^b  - ?D_c^a?\tD_*^{bc}  - D_*^{ca}?\tD_c^b? - y(\rd)  \Lambda D^{ab} \right] \delta g_{ab} \cr
        & -4\alpha_0 K^{bmn} \delta (-\Gamma^c{}_{m(n}D_{b)c}) - 2 D_*^{mn} \nabla_n \delta\Gamma^c{}_{cm} + 2 D_*^{mn}\nabla_c \delta \Gamma^c{}_{nm}~~~.
        \end{split}\nonumber
\end{align}
The variation has split into $\delta g_{ab}$ terms and $\delta \Gamma^c_{mn}$ terms as in Eq.~(\ref{e:varSGamma}). We thus peel off part of the $\Theta^{ab}_{(g)}$ piece, defining
\begin{align}
\begin{split}
        (J_0 c)^{-1}\Theta^{ab}_{(g1)} =& \tfrac{\alpha_0 }{2} \left[  K^{(a}{}_{mn} K^{b)mn}  + 2 K^{mn(a} K_{mn}{}^{b)}\right] \cr
        & - \tfrac{1}{2} ?D_c^(a?\tD_*^{b)c}  - \tfrac{1}{2} D_*^{c(a}?\tD_c^b)? - y(\rd) \Lambda D^{ab}
        \end{split}
\end{align}

Along with this, we integrate by parts in the last two terms and simplify
\begin{align}
        -\tfrac{2}{J_0 c}\delta \mathcal{L}_D =& (J_0 c)^{-1} \Theta^{ab}_{(g1)} \delta g_{ab} + 4 \alpha_0 K^{bmn}D_{c(b}\delta \Gamma^c{}_{n)m} + 2 (\nabla_n D_*^{mn}) \delta \Gamma^c{}_{cm} - 2 \nabla_c D_*^{mn}\delta \Gamma^c{}_{mn}~~~\cr
        =& (J_0 c)^{-1} \Theta^{ab}_{(g1)} \delta g_{ab} + 4 \alpha_0 K^{bmn}D_{cn} \delta\Gamma^c{}_{bm} + 2 (\nabla_n D_*^{mn}) \delta_c{}^b \delta\Gamma^c{}_{bm} \cr
        &- 2 (\nabla_c D_*^{ab}) \delta \Gamma^c_{ab} + 4 \alpha_0 K^{bmn}D_{cb} \delta\Gamma^c{}_{nm}~~~ \cr
        =&  (J_0 c)^{-1} \Theta^{ab}_{(g1)} \delta g_{ab} + \left(4 \alpha_0 K^{abn} D_{cn} + 2 (\nabla_n D_*^{an}) \delta_c{}^b  - 2 \nabla_c D_*^{ab}\right) \delta \Gamma^c{}_{ab}
\end{align}
Plugging this back into the action yields
\begin{align}
        \delta S_D = - \tfrac{1}{2} \int d^{\rd} x \sqrt{|g|} \left[\theta_{(g1)}^{ab} - g^{ab} \mathcal{L}_D \right] \delta g_{ab} + \int d^{\rd} x \sqrt{|g|} \tV_c{}^{ab} \delta \Gamma^c{}_{ab}
\end{align}
where
\begin{align}
        \tV^{cab} = (- J_0 c) \left( \alpha_0 K^{(ab)m}?D_m^c? + \tfrac{1}{2} g^{c(a} \nabla_m D_*^{b)m} - \nabla^c D_*^{ab}\right)~~~.
\end{align}
Comparing with Eq.~(\ref{e:varSGamma}) and reinserting the contribution from the Gauss-Bonnet portion of the source action (which is zero in $\rd = 4$) we find that the stress-energy tensor is
\begin{align}
\begin{split}
        \Theta^{ab} =& \nabla_c\tV^{(ab)c} - \nabla_c \tV^{cab} + \tfrac{\alpha_0 J_0 c}{2} \left[?K^(a_mn? ?K^b)mn? + 2 ?K^cm(a? ?K_cm^b)? \right] \cr 
&- \tfrac{J_0 c}{2} ?D_c^(a? \tD_*^{b)c} -\tfrac{J_0 c}{2} D_*^{c(a} \tD?{}_c^b)? - J_0 c y(\rd)\Lambda D^{ab} -  g^{ab} \mathcal{L}_D + \Theta_{(GB)}^{ab} ~~~.
\end{split}
\end{align}
Defining $V^{cab}$ without the tilde as a scaled version of $\tV^{cab}$ by removing the proportionality factor $-J_0c$
\begin{align}
        V^{cab} =&\alpha_0 K^{(ab)m}?D_m^c? + \tfrac{1}{2} g^{c(a} \nabla_m D_*^{b)m} - \nabla^c D_*^{ab}
\end{align}
the stress-energy tensor can be expressed as
\begin{align}
\begin{split}
        \Theta^{ab} =& J_0 c \left(\nabla_c V^{cab} -\nabla_c V^{(ab)c}\right)  + \tfrac{\alpha_0 J_0 c}{2} \left[?K^(a_mn? ?K^b)mn? + 2 ?K^cm(a? ?K_cm^b)? \right] \cr 
&- \tfrac{J_0 c}{2} ?D_c^(a? \tD_*^{b)c} -\tfrac{J_0 c}{2} D_*^{c(a} \tD?{}_c^b)? - J_0 c y(\rd)\Lambda D^{ab} -  g^{ab} \mathcal{L}_D + \Theta_{(GB)}^{ab}~~~.
\end{split} \label{e:StressEnergyE}
\end{align}

\subsection{Proof of Eq.~(\ref{e:varSGamma})}
We shall find the following useful in this endeavor
\begin{align}
        \label{e:useful3}
        (\sqrt{|g|} T^{abc}),_{c} =& \sqrt{|g|} \left( \nabla_c T^{abc} - \Gamma^a{}_{dc} T^{dbc} - \Gamma^b{}_{dc} T^{adc} \right) \\
        \label{e:useful4}
        (\sqrt{|g|} T^{cab}),_{c} =& \sqrt{|g|} \left( \nabla_c T^{cab} - \Gamma^a{}_{dc} T^{cdb} - \Gamma^b{}_{dc} T^{cad} \right) 
\end{align}
for some tensor $T^{abc}$ and the commas denote partial derivatives
\begin{align}
   (\sqrt{|g|} T^{abc}),_{c}  = \partial_c(\sqrt{|g|} T^{abc})~~~.
\end{align}
We first prove the above useful equations:
\begin{align}\begin{split}
        (\sqrt{|g|} T^{abc}),_{c} =& \sqrt{|g|}_{,c} T^{abc} + \sqrt{|g|} T^{abc}{},_{c} \cr
        =& \tfrac{1}{2} \sqrt{|g|} g^{de}g_{de},_c T^{abc} + \sqrt{|g|} T^{abc},_c \cr
        =& \sqrt{|g|} \left( \Gamma^d{}_{cd} T^{abc} + T^{abc},_c \right)
        \end{split}
\end{align}
Using the relationship between covariant derivatives and ordinary partial derivatives on the last line reproduces the result in Eq.~(\ref{e:useful3}). Eq.~(\ref{e:useful4}) follows by permuting indices.

Now we will prove Eq.~(\ref{e:varSGamma}). We start by expanding out the variation $\delta \Gamma^c{}_{ab}$
\begin{align*}
        \int d^{\rd} x\sqrt{|g|} \tV_c{}^{ab} \delta \Gamma^c{}_{ab} =& \int d^{\rd} x\sqrt{|g|} \tV_c{}^{ab} \left[ \tfrac{1}{2} \delta g^{cd} (g_{d(a , b)} - g_{ab,d}) + \tfrac{1}{2} g^{cd} (2 \delta g_{da,b} - \delta g_{ab,d})\right] 
\end{align*}
Next, we relabel some indices, integrate by parts, and simplify
\begin{align*}
        \int d^{\rd} x\sqrt{|g|} \tV_c{}^{ab} \delta \Gamma^c{}_{ab} =& \int d^{\rd} x\sqrt{|g|}\tV_c{}^{ab} \tfrac{1}{2}   (g_{d(a , b)} - g_{ab,d}) \delta g^{cd} \cr
        & - \int d^{\rd} x \left( \sqrt{|g|} \tV^{dab} \right)_{,b} \delta g_{ad} +  \int d^{\rd} x \tfrac{1}{2}\left( \sqrt{|g|} \tV^{dab} \right)_{,d} \delta g_{ab}  \cr
        =&  -\int d^{\rd} x\sqrt{|g|} \tV^{cab} \tfrac{1}{2} g^{ed} (g_{d(a,b)} - g_{ab,d}) \delta g_{ce} \cr
        & - \int d^{\rd} x \left[ \left( \sqrt{|g|} \tV^{bad} \right)_{,d}   - \tfrac{1}{2} \left( \sqrt{|g|} \tV^{dab} \right)_{,d} \right] \delta g_{ab}
\end{align*}
Now we use the definition of the Christoffel symbol in the first term and Eqs.~(\ref{e:useful3}) and~(\ref{e:useful4}) in the second line
\begin{align}\begin{split}
        \int d^{\rd} x\sqrt{|g|} \tV_c{}^{ab} \delta \Gamma^c{}_{ab} =& -\int d^{\rd} x\sqrt{|g|}\tV^{cab} \Gamma^e{}_{ab} \delta g_{ce}  \cr
        &- \int d^{\rd} x \sqrt{|g|}  \left[ \nabla_d \tV^{bad} - \Gamma^b{}_{ed} \tV^{ead} - \Gamma^a{}_{ed} \tV^{bed} \right. \cr
        & \left. - \tfrac{1}{2} \nabla_d \tV^{dab} + \tfrac{1}{2} \Gamma^a{}_{ed} \tV^{deb} + \tfrac{1}{2} \Gamma^b{}_{ed} \tV^{dae} \right] \delta g_{ab}  \cr
        =&  -\int d^{\rd} x\sqrt{|g|} \left[ \Gamma^a{}_{ed} \tV^{bed} + \tfrac{1}{2} \nabla_c \tV^{(ab)c} - \tfrac{1}{2} \nabla_c \tV^{cab} \right. \cr
        & \left. \hspace*{40 pt} - \Gamma^a{}_{ed} \tV^{deb} - \Gamma^a{}_{ed} \tV^{bed} + \Gamma^a{}_{ed} \tV^{deb} \right] \delta g_{ab}
        \end{split}
\end{align}
The first and fifth terms cancel and the fourth and sixth terms cancel, leaving us with
\begin{align}
        \int d^{\rd} x\sqrt{|g|} \tV_c{}^{ab} \delta \Gamma^c{}_{ab} =&  -\tfrac{1}{2}\int d^{\rd} x\sqrt{|g|} \left[  \nabla_c \tV^{(ab)c} -  \nabla_c \tV^{cab} \right] \delta g_{ab}
\end{align}
Plugging this into the second line of Eq.~(\ref{e:varSGamma}) reduces it to the third line.

\section{Covariant Conservation of the Stress-Energy Tensor}
In this section, we show that the divergence of the stress-energy tensor vanishes. This will require use of the equations of motion
\begin{align}\label{e:EOMa}
        \alpha_0 \nabla_n K^{(ab)n}
        =& -K_*^{ab}~~~.
\end{align}
The stress-energy tensor $\Theta^{ab}$ is
\begin{align}\label{e:Thetaa}
\begin{split}
\Theta^{ab} = & \Theta^{ab}_{(\Gamma)} + \tfrac{\alpha_0 J_0 c}{2} \left[?K^(a_mn? ?K^b)mn? + 2 ?K^cm(a? ?K_cm^b)? \right] \cr 
&- \tfrac{J_0 c}{2} ?D_c^(a? \tD_*^{b)c} -\tfrac{J_0 c}{2} D_*^{c(a} \tD?{}_c^b)? - J_0 c y(\rd)\Lambda D^{ab} -  g^{ab} \mathcal{L}_D
\end{split}
\\
\label{e:ThetaGammaa}
\Theta^{ab}_{(\Gamma)} =&  J_0 c \left(\nabla_c V^{cab} -\nabla_c V^{(ab)c}\right)
\end{align}

The divergence of the stress-energy tensor is
\begin{align}\label{e:DivTheta0}
        \nabla_a \Theta^{ab} =& \nabla_a \Theta^{ab}_{(\Gamma)} + \tfrac{\alpha_0J_0 c}{2} \nabla_a \left(?K^(a_mn? ?K^b)mn? + 2 ?K^cm(a? ?K_cm^b)? \right)  - \nabla^b \mathcal{L}_D \cr
        &-\tfrac{J_0 c}{2}\nabla_a \left(  ?D_c^(a? ?\tD_*^b)c? + ?D_*^c(a? \tD?{}_c^b)? \right)- J_0 c y(\rd) \Lambda \nabla_a D^{ab} 
\end{align}

First, we focus on the $\nabla_a \Theta^{ab}_{(\Gamma)}$ term:  
\begin{align}\label{e:DivThetaGamma}
        -(J_0 c)^{-1}\nabla_a \Theta^{ab}_{(\Gamma)} =& \nabla_a \nabla_c V^{abc} + \nabla_a \nabla_c V^{bac} - \nabla_a \nabla_c V^{cab} \cr
        =& \nabla_a \nabla_c V^{abc} + \nabla_a \nabla_c V^{bac} - \nabla_c \nabla_a V^{acb} \cr
        =& [\nabla_a ,\nabla_c] V^{abc} + \nabla_a \nabla_c V^{bac} \cr
        =& ?R^a_dac? V^{dbc} + ?R^b_dac?V^{adc} + ?R^c_dac? V^{abd} + \nabla_a \nabla_c V^{bac} \cr
        =& -R_{dc}V^{dbc} + ?R^b_dac?V^{adc} + R_{da}V^{abd} + \nabla_a \nabla_c V^{bac} \cr
        =& -R_{dc}V^{dcb}  + ?R^b_dac?V^{adc} + R_{da} V^{adb} + \nabla_a \nabla_c V^{bac} \cr
        =& ?R^b_dac?V^{adc}+ \nabla_a \nabla_c V^{bac} .
\end{align}
In going from the third to fourth line, we have used the following property of the Riemann curvature tensor as pertains to rank three contravariant tensors:
\begin{align}
         [\nabla_e ,\nabla_f] V^{abc}  = ?R^a_def? V^{dbc} + ?R^b_def?V^{adc} + ?R^c_def? V^{abd}
\end{align}
Then Eq.~(\ref{e:DivTheta0}) becomes,
\begin{align}\label{e:DivTheta1}
        \nabla_a \Theta^{ab}  =& \nabla_a \Theta^{ab}_{(\alpha_0)} 
        +J_0 c ?R^b_dac? \left(\nabla^a D_*^{dc}-\tfrac{1}{2} g^{a(c} \nabla_m D_*^{d)m} \right)\cr
        & +J_0 c  \nabla_a \nabla_c \left( \nabla^b D_*^{ac} -\tfrac{1}{2} g^{b(a} \nabla_m D_*^{c)m} \right) 
        -\tfrac{J_0 c}{2} \nabla_a \left( ?D_c^(a? \tD_*^{b)c} + ?\tD_c^(a? D_*^{b)c}\right)  \cr
        &+\tfrac{ J_0 c}{2} \nabla^b \left(  D_{mn}\tD_*^{mn} + y(\rd)\Lambda D\right) - J_0 c y(\rd)\Lambda\nabla_a D^{ab}
\end{align}
where we have collected all terms proportional to $\alpha_0$ into
\begin{align}\label{e:Thetal0}
        \nabla_a \Theta^{ab}_{(\alpha_0)}  =& \nabla_a \Theta^{ab}_{(\Gamma,\alpha_0)} + \tfrac{ \alpha_0 J_0 c }{2} \nabla_a\left(?K^(a_mn? ?K^b)mn? + 2 ?K^cm(a? ?K_cm^b)? \right) \cr
        & -  \tfrac{\alpha_0 J_0 c}{2} \nabla^b( K_{cma}K^{cma}) \\
        \label{e:tglProof0}
        \nabla_a \Theta^{ab}_{(\Gamma,\alpha_0)} = & -\alpha_0 J_0 c \left[?R^b_dac?  K^{(dc)m}?D_m^a? + \nabla_a \nabla_c \left(  K^{(ac)m} ?D_m^b? \right) \right]
\end{align}
Our first goal is to demonstrate that $\nabla_a \Theta^{ab}_{(\Gamma,\alpha_0)}$ can be written independent of $\alpha_0$. We will do so by simplifying it to terms involving only $\nabla_a K^{(cm)a}$ which, via the equation of motion~(\ref{e:EOMa}), simplify to terms independent of $\alpha_0$.  In the following, we will often commute partial derivatives at the cost of generating Riemann curvature tensor terms according to  Eq.~(\ref{e:RactionGeneral}) and simplify via use of the following useful identities
\begin{align}
\nabla_a D_{bm} =& - \frac{1}{2} K_{(bm)a} + \frac{1}{2} \nabla_{(b} D_{m)a} =  \frac{1}{2} K_{mab} + \frac{1}{2} \nabla_{(a} D_{b) m} \\
 K^{amn} \nabla_m D_{nb} =& \frac{1}{2} K^{amn} K_{bmn} \\
 \nabla_b \nabla_m D_{na} =& \nabla_m \nabla_b D_{na} - D_{c(a}R^c{}_{n) bm} \\
 K^{cma} =& \frac{1}{2} K^{(cm)a} - \frac{1}{2} K^{acm} \\
 K^{[cm]a} =& - K^{acm} \\
 ?R^b_[dac]? = 0
\end{align}
Moving common proportional factors in Eq.~(\ref{e:tglProof0}) to the left hand side and simplifying, we have
\begin{align}\label{e:tglProof1}
        \frac{-1}{\alpha_0J_0 c} \nabla_a \Theta^{ab}_{(\Gamma,\alpha_0)} = 
        & ?R^b_dac?  K^{(dc)m} ?D_m^a? + \nabla_a \nabla_c (K^{(ac)m} ?D_m^b?) \cr
        =& ?R^b_dac?  K^{(dc)m} ?D_m^a? + (\nabla_a \nabla_c K^{(ac)m}) ?D_m^b? + K^{(ac)m} \nabla_a \nabla_c ?D_m^b? \cr
        &+ 2 (\nabla_a K^{(ac)m}) \nabla_c ?D_m^b? \cr
        =& ?R^b_dac?  K^{(dc)m} ?D_m^a? +  \{ \nabla_a , \nabla_c \} K^{acm} ?D_m^b?  \cr
        &+ K^{acm} \nabla_a \nabla_c ?D_m^b? + K^{cam} \nabla_a \nabla_c ?D_m^b? \cr
        &+ 2 (\nabla_a K^{acm}) \nabla_c ?D_m^b? + 2 (\nabla_a K^{cam}) \nabla_c ?D_m^b? 
\end{align}
To simplify further, we use the following identity
\begin{align}
        \label{e:Kidentity}
        \{ \nabla_a , \nabla_c \} K^{acm} =& - 2 \nabla_c \nabla_a K^{(cm)a} - 3 ?R^m_dca?K^{cad}
\end{align}
Substituting these into Eq.~(\ref{e:tglProof1}) and simplifying results in
\begin{align}\label{e:tglProof2}
                \frac{-1}{\alpha_0 J_0 c} \nabla_a \Theta^{ab}_{(\Gamma,\alpha_0)} = 
        & ?R^b_dac?  K^{(dc)m} ?D_m^a?  + (- 2\nabla_c \nabla_a K^{(cm)a} - 3 ?R^m_dca?K^{cad})?D_m^b?  \cr
        &+\frac{1}{2} K^{acm} \nabla_a \nabla_{[c} D_{m]}{}^b + K^{cam} \left(\frac{1}{2}\nabla_c \nabla_{[a} ?D_m]^b? - ?R^d_mac? ?D_d^b? + ?R^b_dac? ?D_m^d? \right) \cr
        &+ \nabla_a K^{acm} \nabla_{[c} D_{m]}{}^b - (\nabla_a K^{(cm)a}) \nabla_c ?D_m^b?- (\nabla_a K^{cma}) \nabla_{[c} ?D_m]^b?
\end{align}
In the last terms we have used the antisymmetry of $K^{acm}$, Eq.~(\ref{e:K3}), along with the following property of the Riemann curvature tensor as applies to rank-two tensors of mixed indices:
\begin{align}
        [\nabla_c ,\nabla_a] ?D_m^b? = - ?R^d_mac? ?D_d^b? + ?R^b_dac? ?D_m^d?
\end{align}
Collecting the Riemann curvature terms together and simplifying and substituting $?K_cm^b? = \nabla_{[c} ?D_m]^b?$ results in 
\begin{align}\label{e:tglProof3}
                \frac{-1}{\alpha_0 J_0 c } \nabla_a \Theta^{ab}_{(\Gamma,\alpha_0)} = 
        & 2 ?R^b_dac?K^{cdm} ?D_m^a? - 2 ?R^m_dca? K^{cad} ?D_m^b? \cr
        &- 2 (\nabla_c \nabla_a K^{(cm)a}) ?D_m^b? - (\nabla_a K^{(cm)a} ) \nabla_c ?D_m^b? \cr
        &+ \frac{1}{2}\nabla_a ( ?K^(a_cm? ?K^b)cm?) - (\nabla_a K^{cma} ) ?K^b_cm?
\end{align}
The first five terms are of the form we seek. We are left to simplify the last term:
\begin{align}\label{e:tglProof4}
         - (\nabla_a K^{cma} ) ?K^b_cm? =& - (\nabla_a K^{cma})(-K_{[cm]}{}^b) \cr
         =& (\nabla_a K^{cma}) K_{cm}{}^b - (\nabla_a K^{cma}) K_{mc}{}^b \cr
         =& \nabla_a (K^{cma} K_{cm}{}^b) - K^{cma} \nabla_a K_{cm}{}^b - (\nabla_a K^{cma}) K_{mc}{}^b \cr
         =& 2 \nabla_a (K^{cma}K_{cm}{}^b) - \nabla_a (K^{cma}K_{cm}{}^b) - K^{cma}\nabla_a K_{cm}{}^b - (\nabla_a K^{cma})K_{mc}{}^b \cr
         =& \nabla_a (K^{cm(a}K_{cm}{}^{b)}) - (\nabla_a K^{(cm)a})K_{cm}{}^b - 2 K^{cma}\nabla_a K_{cm}{}^b 
\end{align}
Plugging this back into Eq.~(\ref{e:tglProof3}) yields
\begin{align}\label{e:tglProof5}
        \frac{-1}{\alpha_0 J_0 c } \nabla_a \Theta^{ab}_{(\Gamma,\alpha_0)} = 
        & 2 ?R^b_dac?K^{cdm} ?D_m^a? - 2 ?R^m_dca? K^{cad} ?D_m^b? \cr
        &- 2 (\nabla_c \nabla_a K^{(cm)a}) ?D_m^b? - (\nabla_a K^{(cm)a} ) \nabla_c ?D_m^b? \cr
        &+ \frac{1}{2}\nabla_a ( ?K^(a_cm? ?K^b)cm?) + \nabla_a (K^{cm(a}K_{cm}{}^{b)}) \cr
        & - (\nabla_a K^{(cm)a})K_{cm}{}^b - 2 K^{cma}\nabla_a K_{cm}{}^b 
\end{align}
The third line can be seen to cancel with the other terms in $\nabla_a \Theta^{ab}$ proportional to $\alpha_0$ and quadratic in $K^{cma}$ except for the term proportional to the metric $g^{ab}$. It must be that the very last term in Eq.~(\ref{e:tglProof4}) cancels with the remaining metric term. We now show this:
\begin{align}\label{e:tglProof6}
        - 2 K^{cma}\nabla_a K_{cm}{}^b  = & - 2 K^{cma}\nabla_a (\nabla_m ?D^b_c? - \nabla^b D_{mc}) \cr
        =& - K^{cma}[ \nabla_a , \nabla_m ]?D^b_c? + 2 K^{cma}\nabla_a \nabla^b D_{mc} \cr
        =& - K^{cma} ?R^b_dam? ?D^d_c? +  K^{cma} ?R^d_cam? ?D^b_d? \cr
        &+ 2 K^{cma}\nabla^b \nabla_a  D_{mc} - 2 K^{cma} ?R^d_(c|a|^b? ?D_m)d? \cr
        =& - K^{cmd} ?R^b_adm? ?D^a_c? +  K^{cad} ?R^m_cda? ?D^b_m? \cr
        &+ K^{cma}\nabla^b K_{cam} + 2 K^{cma} ?R^b_ad(c? ?D_m)^a? \cr
        =& -\frac{1}{2}\nabla^b (K^{cma}K_{cma}) + K^{cad} ?R^m_cda? ?D^b_m? \cr
        & - K^{cmd}?R^b_adm? ?D^a_c? + 2 K^{cmd} ?R^b_da(c? ?D_m)^a?
\end{align}
Upon substituting this into Eq.~(\ref{e:tglProof5}), all terms involving the Reimann curvature tensor cancel owing to the Bianchi identity $?R^m_[dac]? = 0$. The remaining terms are
\begin{align}\label{e:tglProof7}
        \frac{-1}{\alpha_0 J_0 c } \nabla_a \Theta^{ab}_{(\Gamma,\alpha_0)} = 
        &- 2 (\nabla_c \nabla_a K^{(cm)a}) ?D_m^b? - (\nabla_a K^{(cm)a} ) \nabla_c ?D_m^b? \cr
        &+ \frac{1}{2}\nabla_a ( ?K^(a_cm? ?K^b)cm?) + \nabla_a (K^{cm(a}K_{cm}{}^{b)}) \cr
        & - (\nabla_a K^{(cm)a})K_{cm}{}^b -\frac{1}{2}\nabla^b (K^{cma}K_{cma})  
\end{align}
Upon substituting this into Eq.~(\ref{e:Thetal0}), all terms cancel aside from three terms related to the equations of motion
\begin{align}
        (J_0c)^{-1}\nabla_a \Theta^{ab}_{(\alpha_0)} = & 2\alpha_0 (\nabla_c \nabla_a K^{(cm)a})?D_m^b? +
         \alpha_0(\nabla_a K^{(cm)a}) \nabla_c ?D_m^b?  \cr
         &+ \alpha_0 (\nabla_a K^{(cm)a})?K_cm^b?
\end{align}
These three terms can be combined into two terms
\begin{align}
        (J_0c)^{-1}\nabla_a \Theta^{ab}_{(\alpha_0)} = & 2\alpha_0 \nabla_c (?D_m^b? \nabla_a K^{(cm)a}) - \alpha_0(\nabla_a K^{(cm)a}) \nabla^b D_{cm}
\end{align}
Now we can substitute in the equations of motion~(\ref{e:EOMa}) to remove the $\alpha_0$ dependence
\begin{align}\label{e:divThetaLambdaPenultimate}
        (J_0c)^{-1}\nabla_a \Theta^{ab}_{(\alpha_0)} =&-2\nabla_c (?D_a^b? K_*^{ca}) + K_*^{ac}\nabla^b D_{ac}    
\end{align}
Using Eqs.~(\ref{e:tDab}) and~(\ref{e:Star}) we can show
\begin{align}\label{e:Kstar}
        K_*^{ca} =& - R_*^{ca} - \tD_*^{ca} - \tfrac{1}{2} y(\rd) \Lambda g^{ca}
\end{align} 
Using this equation along with some rearrangement of the derivatives, we can expand the last term in Eq.~(\ref{e:divThetaLambdaPenultimate}) as  follows:
\begin{align}
        K_*^{ac}\nabla^b D_{ac}    =& -\tfrac{1}{2} \nabla^b (D_{ac} \tD_*^{ac}) - D_*^{ac} \nabla^b R_{ac} - \tfrac{1}{2} y(\rd) \Lambda \nabla^b D
\end{align}
With this expansion, Eq.~(\ref{e:divThetaLambdaPenultimate}) becomes
\begin{align}\label{e:divThetaLambdafinal}
        (J_0c)^{-1}\nabla_a \Theta^{ab}_{(\alpha_0)} =&-\tfrac{ 1}{2} \nabla^b \left(  D_{ac}\tD_*^{ac} + y(\rd)\Lambda D\right)    -2 \nabla_c (?D_a^b? K_*^{ca}) -D_*^{ac}\nabla^b R_{ac}  
\end{align}
Upon substituting into Eq.~(\ref{e:DivTheta1}), the first term of Eq.~(\ref{e:divThetaLambdafinal}) cancels with the first term on the last line of Eq.~(\ref{e:DivTheta1}). After some simplifications, this leaves us with the following for the divergence of the full stress energy tensor
\begin{align}\label{e:DivTheta2}
\begin{split}
        \nabla_a \Theta^{ab}\propto & -2  \nabla_c (?D_a^b? K_*^{ca}) - D_*^{ac}\nabla^b R_{ac} + ?R^b_dac? \left[\nabla^a D_*^{dc} - \frac{1}{2} g^{a(c} \nabla_m D_*^{d)m} \right] \cr
        & +\nabla_a \nabla_c \left[  \nabla^b D_*^{ac}  - \frac{1}{2} g^{b(a} \nabla_m D_*^{c)m} \right] \cr
        & - \frac{1}{2} \nabla_a \left[ ?D_c^(a? \tD_*^{b)c} +  ?\tD_c^(a? D_*^{b)c} + 2 y(\rd) \Lambda D^{ab}\right]
        \end{split}
\end{align}
with the proportionality constant equal to $J_0c$.
Next, we expand and substitute in for $\tD_{ab}$ leaving us with
\begin{align}\label{e:DivTheta3}
        \nabla_a \Theta^{ab}  \propto  &-2  \nabla_c (?D_a^b? K_*^{ca})  - D_*^{ac}\nabla^b R_{ac}  
        +  ?R^b_dac? \nabla^a D_*^{dc} - \frac{1}{2} ?R_c^b? \nabla_a D_*^{ca}\cr
        & +\left[\nabla_a \nabla_c\nabla^b - \frac{1}{2}\nabla_a \nabla^b \nabla_c - \frac{1}{2}\nabla^b\nabla_a  \nabla_c \right] D_*^{ac} \\
        &       - \frac{1}{2} \nabla_a \left[ 2(\rd-1)?D_c^(a? D_*^{b)c}  - 2 ?D_c^(a? R_*^{b)c} + y(\rd) \Lambda ?D_c^(a?g^{b)c}  - 2 ?R_c^(a? D_*^{b)c} \right] \nonumber
\end{align}
Using again Eq.~(\ref{e:Kstar}), the first three terms in the last line can be collapsed into a single term involving $K_*^{bc}$. Doing this as well as reorganizing the triple derivative terms leaves us with
\begin{align}\label{e:DivTheta4}
\begin{split}
        \nabla_a \Theta^{ab} \propto & -2  \nabla_c (?D_a^b? K_*^{ca}) - D_*^{ac}\nabla^b R_{ac}  
        +  ?R^b_dac? \nabla^a D_*^{dc} - \frac{1}{2} ?R_c^b? \nabla_a D_*^{ca}\cr
        &  +\left[\nabla_a [\nabla_c , \nabla^b] - \frac{1}{2} [\nabla^b , \nabla_a ] \nabla_c  \right] D_*^{ac}  + \nabla_a(?D_c^(a? K_*^{b)c})  + \nabla_a (?R_c^(a?D_*^{b)c}  )
\end{split}
\end{align}
Substituting the Riemann and Ricci tensors in the commutator terms and simplifying and combining the first terms on the first line with the last two terms on the last line results in
\begin{align}\label{e:DivTheta5}
\begin{split}
        \nabla_a \Theta^{ab}  \propto & \nabla_a (?D_c^[a? K_*^{b]c}) - D_*^{ac}\nabla^b R_{ac}  
        +  ?R^b_dac? \nabla^a D_*^{dc} - \frac{1}{2} ?R_c^b? \nabla_a D_*^{ca}  \cr
        & - D_*^{dc} \nabla^a?R^b_dac? - ?R^b_dac? \nabla^aD_*^{dc} - \nabla_a(?R^b_c?D_*^{ac}) - \frac{1}{2}?R^b_d?\nabla_c D_*^{dc}\cr
        &  + \nabla_a (?R_c^a?D_*^{bc}  ) + \nabla_a (?R^b_c?D_*^{ac}  )
        \end{split}
\end{align}
The  third and sixth terms cancel, the seventh and tenth terms cancel, and the fourth and eighth terms combine. These simplifications along with using Eqs.~(\ref{e:tDab}) and (\ref{e:Star}) to expand $K_*^{bc}$ in the first term leads to
\begin{align}\label{e:DivTheta6}
        \nabla_a \Theta^{ab}  \propto & -2 (2 \rd -3) \nabla_a (?D_c^[a? R^{b]c} ) - D_*^{ac}\nabla^b R_{ac}  
        - ?R_c^b? \nabla_a D_*^{ca}  -D^*_{ac} \nabla_d R^{bcda}\cr
        & + \nabla_a (?R_c^a?D_*^{bc}  )
\end{align}
where we have also raised and relabeled indices in the Riemann term. We can use the contraction of the second Bianchi identity to simplify the term with the Riemann tensor 
\begin{align}
        \nabla_d R^{bcda} = \nabla^{c} ?R^ba?-\nabla^{b} ?R^ca?
\end{align}
This leads us to
\begin{align}\label{e:DivTheta7}
        \nabla_a \Theta^{ab}  \propto &-2 (2 \rd -3) \nabla_a (?D_c^[a? R^{b]c} ) - D_*^{ac}\nabla^b R_{ac}  
        - ?R_c^b? \nabla_a D_*^{ca}  \cr
        & -D^*_{ac}\nabla^{c} ?R^ba?+D^*_{ac}\nabla^{b} ?R^ca?+ \nabla_a (?R_c^a?D_*^{bc}  )
\end{align}
The second and fifth terms cancel and we combine the rest into total derivatives and relabel indices leaving us with
\begin{align}\label{e:DivTheta8}
        \nabla_a \Theta^{ab}  \propto & -2 (2 \rd -3) \nabla_a (?D_c^[a? R^{b]c} ) 
        - \nabla_a (D_*^{c[a} ?R_c^b]?) 
\end{align}
Using Eq.~(\ref{e:Star}) to expand the second term leads to
\begin{align}\label{e:DivTheta9}
        \nabla_a \Theta^{ab}  \propto & - 2 (2 \rd -3) \nabla_a (?D^c[a? ?R_c^b]? ) - (\rd -1) \nabla_a (D g^{c[a} ?R_c^b]? )\cr
        &+ 2 (2 \rd - 3) \nabla_a ( D^{c[a} ?R_c^b]?) 
\end{align}
The first and third terms cancel and the second term is zero due to the symmetry of the Ricci tensor. 
Therefore, we have shown that
\begin{align}
        \nabla_a \Theta^{ab} = 0~~~.
\end{align}


\bibliographystyle{JHEP}
\bibliography{Bibliography}
\end{document}